\documentclass[a4paper,fleqn,usenatbib]{mnras}
\usepackage{epsfig}
\usepackage{graphicx,wrapfig,lipsum}	
\usepackage{amsmath}     
\usepackage{hyperref}	    
\usepackage{amssymb}   	
\usepackage{multicol}    
\usepackage{multirow}
\usepackage{bm}		    
\usepackage{pdflscape}

\usepackage[T1]{fontenc}
\usepackage{ae,aecompl}
\usepackage{txfonts}
\usepackage{array}
\usepackage{caption}
\usepackage{longtable}
\usepackage{threeparttable}
\title[$T_{\rm eff}$, radius and luminosity of M-dwarfs]{Estimating $T_{\rm eff}$, radius and luminosity of M-dwarfs using high resolution optical and NIR spectral features}
\author[Khata et al. 2021]{
Dhrimadri Khata,$^{1}$\thanks{E-mail: dhrimadrikht@gmail.com (DK)}
Soumen Mondal,$^{1}$
Ramkrishna Das,$^{1}$
Tapas Baug$^{1}$
\\
$^{1}$Satyendra Nath Bose National Centre for Basic Sciences,
Block-JD, Sector-III, Salt Lake, Kolkata-700 106
}

\date{Accepted XXX. Received YYY; in original form ZZZ}

\pubyear{2021}

\begin{document}
\label{firstpage}
\pagerange{\pageref{firstpage}--\pageref{lastpage}}
\maketitle

\begin{abstract}
We estimate effective temperature ($T_{\rm eff}$), stellar radius, and luminosity for a sample of 271 M-dwarf stars (M0V-M7V) observed as a part of CARMENES (Calar Alto high-Resolution search for M dwarfs with Exo-earths with Near-infrared and optical Echelle Spectrographs) radial-velocity planet survey. For the first time, using the simultaneously observed high resolution (R$\sim90000$) spectra in the optical (0.52\textendash0.96 $\mu$m) and near-infrared (0.96\textendash 1.71 $\mu$m) bands, we derive empirical calibration relationships to estimate the fundamental parameters of these low-mass stars. We select a sample of nearby and bright M-dwarfs as our calibrators for which the physical parameters are acquired from high-precision interferometric measurements. To identify the most suitable indicators of $T_{\rm eff}$, radius, and luminosity (log L/$L_{\sun}$), we inspect a range of spectral features and assess them for reliable correlations. We perform multivariate linear regression and find that the combination of pseudo equivalent widths and equivalent width ratios of the Ca II at 0.854 $\mu$m and Ca II at 0.866 $\mu$m lines in the optical and the Mg I line at 1.57 $\mu$m in the NIR give the best fitting linear functional relations for the stellar parameters with root mean square errors (RMSE) of 99K, 0.06 $R_{\sun}$ and 0.22 dex respectively. We also explore and compare our results with literature values obtained using other different methods for the same sample of M dwarfs.

\end{abstract}



\section{Introduction :}

The properties of planets are directly connected with the fundamental physical properties of the host stars (Santos et al. \citet{Santos2004}; Mann et al. \citet{Mann2013b}). The small size, low mass range (0.075$-$0.50$M_{\sun}$ , Delfosse et al. \citet{Delfosse2000}), presence of protoplanetary disks (Laughlin et al. \citet{Laughlin2004},  Mercer $\&$ Stamatellos \citet{Mercer2020}) and chemically rich environment (Ida $\&$ Lin \citet{Ida2004}) of M-dwarf stars make them prime targets for exoplanet and brown dwarf searches (Bonfils et al. \citet{Bonfils2013}; Anglada-Escud\'e et al. \citet{Anglada2016}). The photometric transit of a planet helps to measure the planet to star radius ratio, and the radial velocity or astrometry wobbles of a star allows to determine the planet to star mass ratio. Though the intrinsic faintness and variation of the properties across all M-spectral subtypes make the task challenging, the precise estimation of the stellar parameters of M dwarfs is extremely important in many aspects of stellar and galactic astronomy: understanding the chemical evolution and star-formation history (Bochanski et al. \citet{Bochanski2007}), tracing the galactic disc kinematics (e.g., Hawley, Gizis $\&$ Reid \citet{Hawley1996}; Reid, Gizis $\&$ Hawley  \citet{Reid2002}), studying stellar age-velocity relations (West et al. \citet{West2006}), constraining galactic structure (e.g., Reid et al. \citet{Reid1997}; Pirzkal et al. \citet{Pirzkal2005}) and the galaxy's mass and luminosity (e.g., Hawkins $\&$ Bessell \citet{Hawkins1988}; Kirkpatrick et al. \citet{Kirkpatrick1994}; Zheng et al. \citet{Zheng2001}, \citet{Zheng2004}).\\ 
$~~~~$ Earlier work constraining the effective temperature ($T_{\rm eff}$) of M dwarfs has been carried out using photometry (Leggett et al. \citet{Leggett1996}) and blackbody fitting techniques (Bessell \citet{Bessell1991}). The straightforward way to derive the $T_{\rm eff}$ is by using the angular diameter estimated from stellar evolution tracks through interferometry (Boyajian et al. \citet{Boyajian2012}), and bolometric flux obtained from multi-band photometry (Nordstr\" om et al. \citet{Nordström2004}). Another method to determine the parameters is by comparing  M dwarf spectra with synthetic models for cool stellar atmospheres (i.e., Casagrande et al. \citet{Casagrande2008}; Gaidos \citet{Gaidos2013}; Dressing $\&$ Charbonneau \citet{Dressing2013}). Veyette et al. (\citet{Veyette2017}) and Rajpurohit et al. (\citet{Rajpurohit2018}) used  advanced BT-Settl models (Allard et al. \citet{Allard2012}, \citet{Allard2013}) to determine the stellar parameters of M dwarfs using CARMENES high resolution spectra (Reiners et al. \citet{Reiners2018}). However, the formation of molecules and dust grains due to cool temperature and inaccurate modeling of the magnetic field induced by the deep convective zones in M dwarf interiors (Mullan $\&$ MacDonald \citet{Mullan2001}; Browning \citet{Browning2008}), constrain the accuracy of theoretical model predictions.\\
$~~~~$ As the direct measurement of the stellar radius is limited to few nearby bright M dwarfs (Berger et al. \citet{Berger2006}), empirical calibrations can be determined and applied to other M dwarfs. Ballard et al. (\citet{Ballard2013}) used interferometric radii to constrain the $T_{\rm eff}$ and radius of Kepler-61b; Boyajian et al. (\citet{Boyajian2012}) calculated radii for K- and M-stars using a mass-radius relation; Johnson et al. (\citet{Johnson2012}) constrained the mass and radius for Kepler Object of Interest (KOI) 254. To derive the metallicity calibration relations for M dwarfs, the use of an F, G, or K binary companion with known metallicity is a well adopted method both in spectroscopy (Newton et al. \citet{Newton2014}; Mann et al. \citet{Mann2013a}) and in photometry (Bonfils et al. \citet{Bonfils2005}; Casagrande et al. \citet{Casagrande2008}; Schlaufman $\&$ Laughlin \citet{Schlaufman2010}). More recently machine learning techniques have been applied widely, where different neural network architectures are built and trained to determine stellar parameters of M dwarfs (Sarro et al. \citet{Sarro2018}, Sharma et al. \citet{Sharma2019}, Antoniadis-Karnavas et al. \citet{Antoniadis2020}, Passegger et al. \citet{Passegger2020}).\\
$~~~~$ The measurement and identification of pseudo equivalent widths (pEWs) and spectral indices of various atomic and molecular features in the optical and NIR bands that are sensitive to different stellar parameters, can also be used as an effective alternative approach to obtain cool star's physical parameters. This method has been widely applied to M dwarfs by Mann, Gaidos $\&$ Ansdell (\citet{Mann2013b}) and Newton et al. (\citet{Newton2015}) to estimate effective temperature, radius, and luminosity. The M dwarf's $T_{\rm eff}$-scale was defined using optical TiO, VO, CrH, FeH equivalent widths by Tinney $\&$ Reid \citet{Tinney1998} and using NIR $H_{2}O$ indices by Delfosse et al. \citet{Delfosse1999}. Rojas-Ayala et al. (\citet{RojasAyala2012}) estimated spectral-type and $T_{\rm eff}$ using the K-band $H_{2}O$  index and metallicity using pEWs of the K-band Na I and Ca I features along with the $H_{2}O$ index. Khata et al. (\citet{Khata2020}) applied the $H_{2}O$-H index defined by Terrien et al. \citet{Terrien2012} in the NIR calibration relations of $T_{\rm eff}$ and radius for M dwarfs and evaluated the spectral type and absolute $K_{s}$ magnitude as a linear function of both H- and K-band $H_{2}O$ indices. Also, Neves et al. (\citet{Neves2012}, \citet{Neves2013}) used pEWs to estimate metallicity and $T_{\rm eff}$ for M dwarfs, and Woolf $\&$ Wallerstein (\citet{Woolf2006}) derived CaH and TiO molecular indices to estimate metallicities of K- and M-dwarfs.\\
$~~~~$ With the development of high-resolution spectrographs, spectra of distant and faint M dwarfs with good signal-to-noise ratio has become available both in the optical (i.e., HARPS, Mayor et al. \citet{Mayor2003}; HARPS-N, Cosentino et al. \citet{Cosentino2012}) and in the NIR wavelength (CRIRES, Kaeufl et al. \citet{Kaeufl2004}). The CARMENES (Reiners et al. \citet{Reiners2018}) radial velocity survey offers an atlas of high resolution (R > 80000) M dwarf spectra abundant with numerous features simultaneously observed in the optical and NIR from 520$-$1710 nm (where M dwarfs emit maximum flux). To minimize trend degeneracies and biases due to M dwarf environment, the simultaneous use of equivalent widths from the optical and NIR is proving very fruitful. As the EWs of the lines with higher excitation potential ($\chi$) change faster with temperature than those with lower excitation potential (Gray \citet{Gray1994}), the ratio of such EWs that have different temperature sensitivity have been extensively used as $T_{\rm eff}$ indicators for solar-type stars (i.e. Gray \citet{Gray1994}; Kovtyukh et al. \citet{Kovtyukh2003}; Montes et al. \citet{Montes2006}; Sousa et al. \citet{Sousa2010}), giant type stars (i.e. Gray \citet{Gray1989}; Strassmeier $\&$ Schordan \citet{Strassmeier2000}) and for super-giants (i.e. Kovtyukh $\&$ Gorlova \citet{Kovtyukh2000}). The benefit of estimating stellar parameters using the EW ratios is that it is independent of the effect of interstellar reddening, spectral resolution, rotational and microturbulence broadening (Kovtyukh et al. \citet{Kovtyukh2003}).\\
$~~~~$ In this paper, we estimate $T_{\rm eff}$, radius and luminosity using sensitive pseudo-EWs and EW ratios from the optical and NIR for a sample of 271 M dwarfs. Though good parallax measurements are available from the Gaia DR2 and Gaia EDR3 catalog (Gaia Collaboration et al. \citet{Gaia2018} $\&$ \citet{Gaia2021}) for our selected CARMENES sample of potential planet-hosting M dwarfs (except one object, 2M J06572616+7405265), the parameters estimated from the spectral features are proven to be strikingly useful. The Spectroscopic data is obtained from the CARMENES radial velocity (RV) survey, which observed 324 M dwarfs during guaranteed time observations (GTO) to search for any orbiting planets.

\section{DATA ACQUISITION $\&$ CALIBRATORS SELECTION:}

We worked with a sample of 271 bright, nearby M-dwarfs with a good signal-to-noise ratio (S/N > 75) across all M-spectral subtypes (M0V-M7V), based on data from the CARMENES data archive at \texttt{CAB (INTA-CSIC)}\footnote{\url{http://carmenes.cab.inta-csic.es/gto/jsp/reinersetal2018.jsp}} (Reiners et al. \citet{Reiners2018}). The spectrograph, installed on the Zeiss 3.5 m telescope, is located at Calar Alto observatory in Spain, and the data was reduced using an automatic pipeline (Zechmeister et al. \citet{Zechmeister2014}) and more details are given in Caballero et al. (\citet{Caballero2016}) and Nidever et al. (\citet{Nidever2015}). CARMENES provides a wide range of high resolution spectra both in the optical channel with wavelength from 0.52\textendash 0.96 $\mu$m and spectral resolution, R $\thicksim$ 94600; and in the NIR channel with wavelength from 0.96\textendash 1.71 $\mu$m and resolution, R $\thicksim$ 80400. After rejecting several double-lined spectroscopic binaries (SB2s), Reiners et al. (\citet{Reiners2018}) came up with an initial sample of 324 M-dwarfs with an average age of 5 Gyr (Cort\'es-Contreras et al. \citet{Cortes2017}; Passegger et al. \citet{Passegger2018}) and a typical limit for the J-band magnitude ($m_{J}$ = 10 mag). From this sample, we selected mainly isolated M dwarfs or those having widely separated (> 5\arcsec) companions and ruled out the fast rotators and stars with signatures of activity that affect the profiles of the spectral lines (Passegger et al. \citet{Passegger2018}; Schweitzer et al. \citet{Schweitzer2019}).  

We choose a sample of 23 target stars with high precision long-baseline interferometric measurements of radii (Boyajian et al. \citet{Boyajian2012}; Mann et al. \citet{Mann2015}; Markus Rabus et al. \citet{Markus2019}) for the calibration of stellar parameters. We preferably select 13 calibrators from Newton et al. (\citet{Newton2015}) where the bolometric luminosities ($L_{bol}$) are measured from multicolor photometry (Boyajian et al. \citet{Boyajian2012}) and effective temperatures are calculated using the interferometric radius and bolometric flux (Mann et al. \citet{Mann2013b}). We supplement this calibration sample with 6 M-dwarfs taken from Mann et al. (\citet{Mann2015}), where the bolometric flux ($F_{bol}$) was calculated by integrating over the radiative flux density and $T_{\rm eff}$ was calculated by fitting the optical spectra with the PHOENIX (BT-Settl) models (Allard et al. \citet{Allard2013}). The remaining 4 calibrators were selected from Markus Rabus et al. (\citet{Markus2019}), where they estimated the $F_{bol}$ using PHOENIX models (Husser et al. \citet{Husser2013}) along with photometric observations and $T_{\rm eff}$ using the Stefan-Boltzmann law. The measured parameter values of our calibration sample cover $T_{\rm eff}$ from 2930 K to 4054 K, stellar radius from 0.183 $R_{\sun}$ to 0.608 $R_{\sun}$ and bolometric luminosity (log L/$L_{\sun}$) from \textendash 2.625 to \textendash 1.048.

\begin{figure*}
\centering
\begin{tabular}{cc}
\hspace*{-0.50cm} \includegraphics[width= 6.5 in, height = 1.96 in, clip]{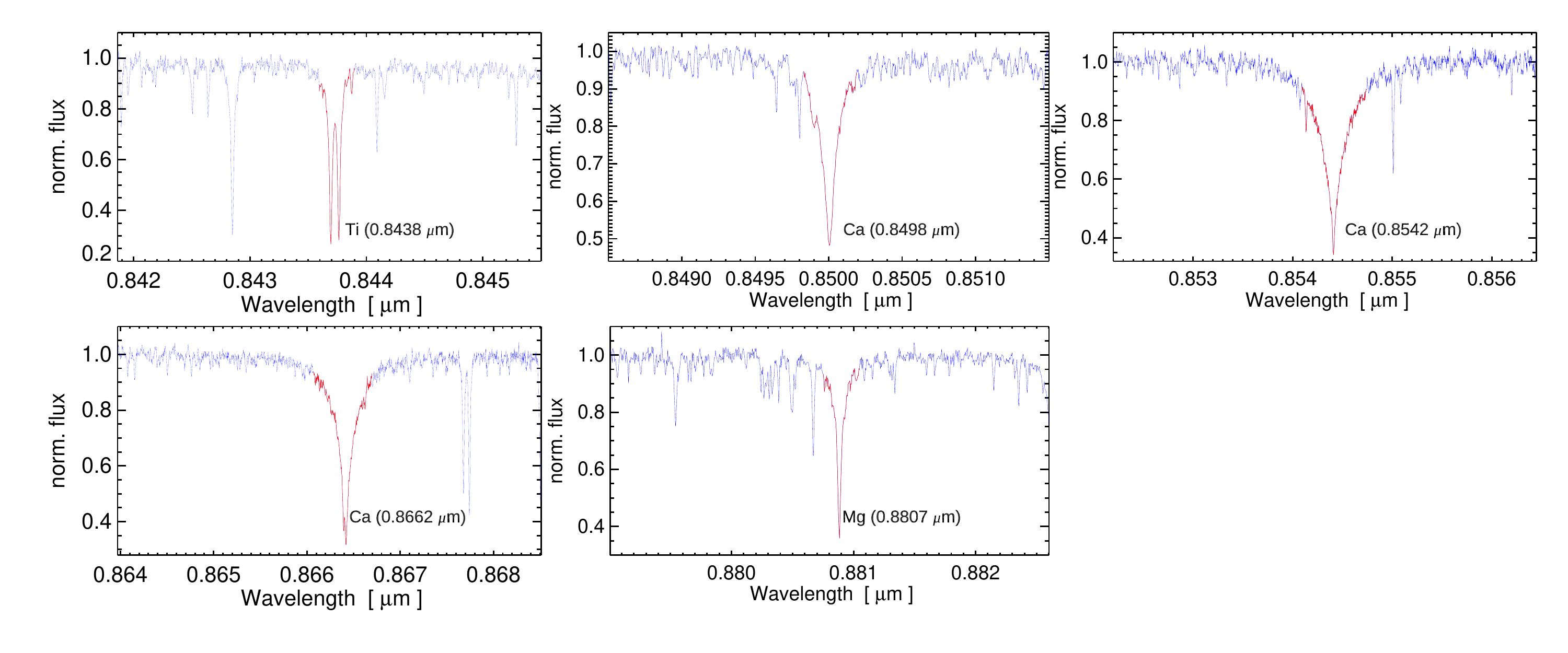} \\
\hspace*{-0.50cm} \includegraphics[width= 6.5 in, height = 1.96 in, clip]{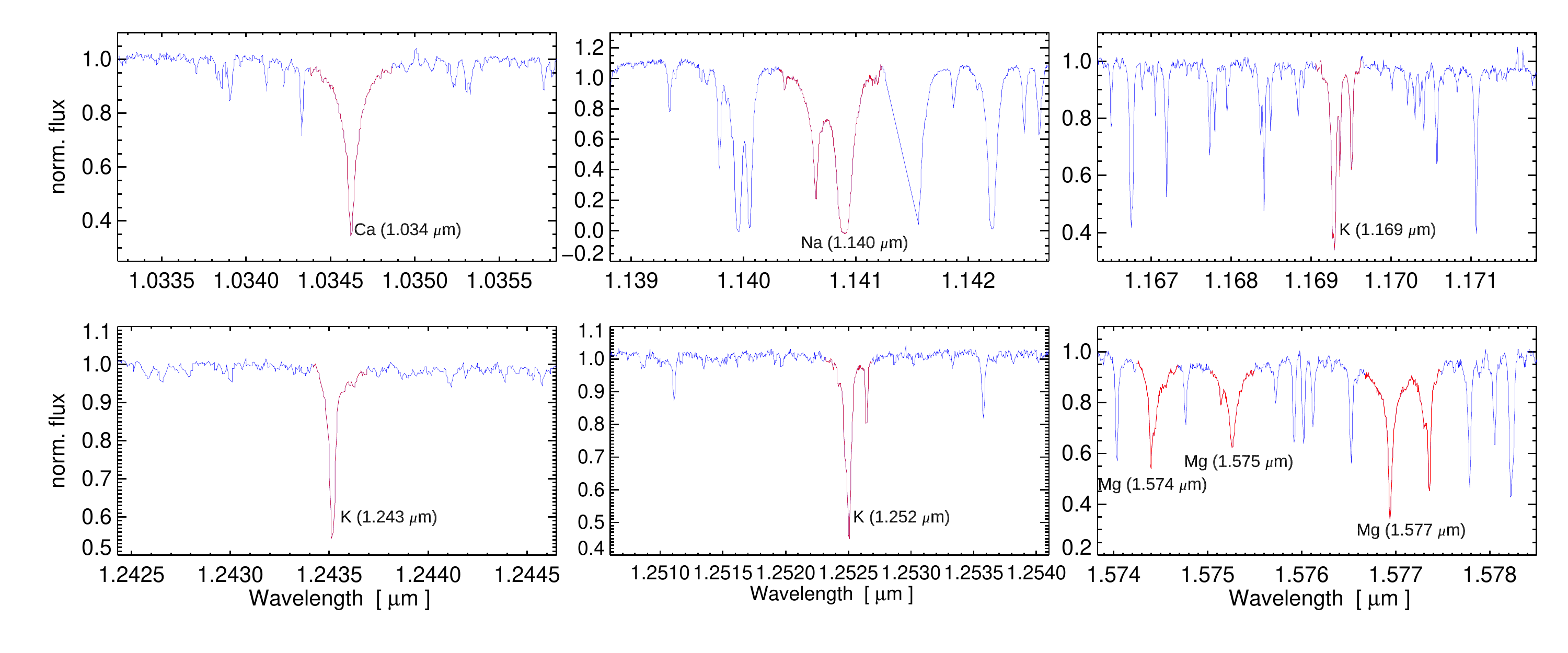}
\end{tabular}
\caption{ A representation of atomic absorption features in optical (Ti I [0.844 $\mu$m], Ca II [0.850 $\mu$m, 0.854 $\mu$m, 0.866 $\mu$m], Mg I [0.881 $\mu$m]) and Near-infrared (Ca I [1.035 $\mu$m], Na I [1.141 $\mu$m], K I [1.169 $\mu$m, 1.240 $\mu$m, 1.250 $\mu$m], Mg I [1.570 $\mu$m]) for the CARMENES spectra of the M-Dwarf GJ 2 (J00051+457) of spectral type M1.0V. We have calculated the EWs of these features with the feature window shown in red color to find and establish the strongest correlations with the stellar parameters.} 
\label{fig:fig1_spectral_features}
\end{figure*}

\section{Result and Discussion :}
\subsection{Spectroscopic Analysis : Optical and NIR EW estimation}

The presence of broad and complex molecular absorption bands in the optical (TiO, VO, CaH) and NIR (H$_{2}$O, FeH, CO, OH) over the entire spectral sequence of M dwarfs, make it difficult to identify the weak atomic lines in the spectra. The atomic features, such as Ti I (0.844 $\mu$m), Ca II (0.850 $\mu$m, 0.854 $\mu$m, 0.866 $\mu$m), Mg I (0.881 $\mu$m) in the optical band and the lines Ca I (1.035 $\mu$m), Na I (1.141 $\mu$m), K I (1.169 $\mu$m, 1.240 $\mu$m, 1.250 $\mu$m), Mg I (1.570 $\mu$m) in the NIR are strong enough and comparatively free from any blends and telluric contaminations. These spectral features are ideal tracers of $T_{\rm eff}$, radius, and luminosity as the line strengths or equivalent widths depend on these stellar parameters (Rajpurohit et al. \citet{Rajpurohit2018}; Newton et al. \citet{Newton2015}). In Fig. \ref{fig:fig1_spectral_features} we show a representation of the absorption lines in the CARMENES spectra of the object GJ 2 (J00051+457) with the feature window colored in red. Some of the NIR atomic absorption features and indices that show significant correlation with stellar properties like Mg I at 1.50 $\mu$m, Al I at 1.67 $\mu$m, Mg I at 1.71 $\mu$m and $H_{2}O$-H index defined from 1.595 $\mu$m to 1.780 $\mu$m (Newton et al. \citet{Newton2015}; Khata et al. \citet{Khata2020}) are absent in the analysis as they are completely or partially in the gap of the CARMENES spectral format. In the J- band region, from 1.20 $\mu$m to 1.48 $\mu$m, the two K I lines at 1.243 $\mu$m and 1.252 $\mu$m stand out prominently, but the rest of this region is very poor in absorption lines due to wide molecular absorptions, telluric $O_{2}$ absorption and strong $H_{2}O$ opacity. We have measured the EWs of the selected lines using the standard equation :

\vspace*{-0.2 cm} \begin{equation} \label{eq1}
\hspace*{2.5cm}  EW_{\lambda} = \int_{\lambda1}^{\lambda2} [1 - \frac{F(\lambda)}{F_{c}(\lambda)}] d\lambda
\end{equation}

\begin{table}
 \centering
 \caption{Optical and NIR spectral features and continuum points}
 \hspace*{-0.3 cm}
 \begin{tabular}{lccc}
  \hline \hline
  Feature & Feature window & Blue Continuum & Red Continuum \\
   & ($\mu$m) & ($\mu$m) & ($\mu$m)\\
  \hline
  
  Ti I (0.844 $\mu$m)$^{a}$ & 0.8436 \hspace*{0.2cm}0.8440 & 0.8431 \hspace*{0.2cm}0.8436 & 0.8443\hspace*{0.2cm} 0.8448 \\
  Ca II (0.850 $\mu$m)$^{a}$ & 0.8498 \hspace*{0.2cm} 0.8504 & 0.8485 \hspace*{0.2cm}0.8496 & 0.8504\hspace*{0.2cm} 0.8512 \\
  Ca II (0.854 $\mu$m)$^{a}$ & 0.8540 \hspace*{0.2cm}0.8550 & 0.8530\hspace*{0.2cm} 0.8539 & 0.8552\hspace*{0.2cm} 0.8560\\  
  Ca II (0.866 $\mu$m)$^{a}$ 	& 0.8660\hspace*{0.2cm} 0.8670	& 0.8650\hspace*{0.2cm} 0.8660	& 0.8670 \hspace*{0.2cm}0.8676	\\ 
  Mg I (0.881 $\mu$m)$^{a}$ & 0.8808 \hspace*{0.2cm}0.8812 & 0.8798 \hspace*{0.2cm} 0.8803 & 0.8815 \hspace*{0.2cm} 0.8822 \\
  Ca I (1.034 $\mu$m)$^{a}$ & 1.0343 \hspace*{0.2cm}1.0350 & 1.0335 \hspace*{0.2cm}1.0341 & 1.0352 \hspace*{0.2cm} 1.0358 \\  
  Na I (1.140 $\mu$m)$^{b}$ & 1.1404 \hspace*{0.2cm}1.1413 & 1.1389 \hspace*{0.2cm}1.1397 & 1.1425 \hspace*{0.2cm} 1.1434 \\
  K I (1.169 $\mu$m)$^{b}$ & 1.1691 \hspace*{0.2cm} 1.1697 & 1.1686 \hspace*{0.2cm} 1.1691 & 1.1697 \hspace*{0.2cm} 1.1705 \\
  K I (1.243 $\mu$m)$^{b}$ & 1.2433 \hspace*{0.2cm} 1.2438 & 1.2427 \hspace*{0.2cm} 1.2433 & 1.2438 \hspace*{0.2cm} 1.2444 \\
  K I (1.252 $\mu$m)$^{b}$ & 1.2523 \hspace*{0.2cm}1.2528 & 1.2513 \hspace*{0.2cm}1.2520 & 1.2529 \hspace*{0.2cm} 1.2535 \\
  Mg I (1.574 $\mu$m)$^{b}$ 	& 1.5740 \hspace*{0.2cm} 1.5780	& 1.5720\hspace*{0.2cm} 1.5740 & 1.5780 \hspace*{0.2cm}1.5800	\\  
      
  \hline
 \end{tabular} 
 \vspace{1ex}
 
  \textbf{Notes.}
     \raggedright \scriptsize \texttt{ Wavelengths are given in vacuum; $^{(a)}$ Atomic features are identified in Reiners et al. (\citet{Reiners2018}), feature and continuum windows are defined based on our observations; $^{(b)}$ Feature and continuum windows are modified from those defined in Newton et al. (\citet{Newton2015})}
 
\label{tab:table1_spectralfeature}
\end{table}

\begin{figure*}
\centering
\vspace*{-0.3 cm}
\begin{tabular}{c}
\hspace*{-0.45cm} \includegraphics[width= 7.2 in, height = 5.6 in, clip]{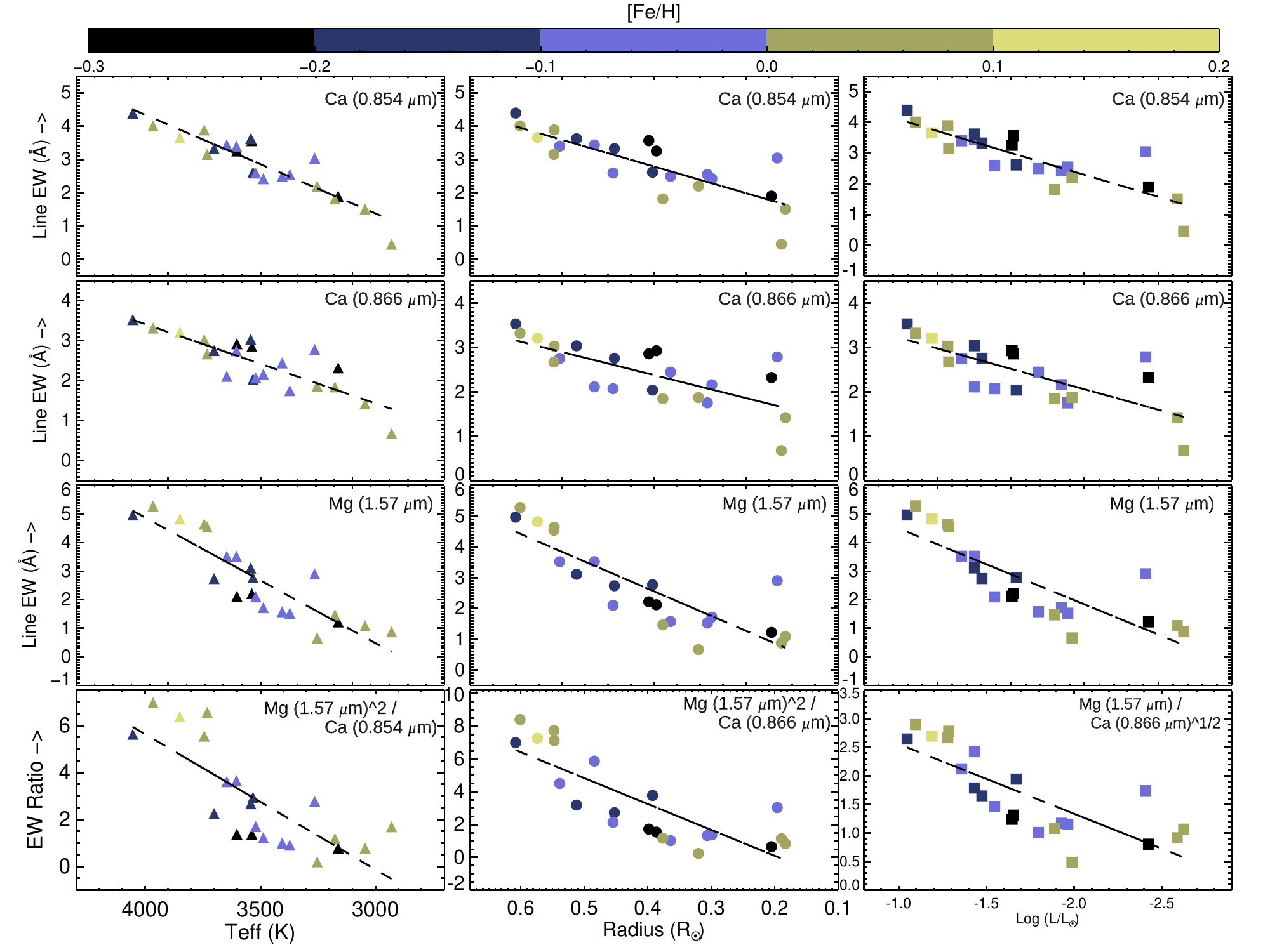}
\end{tabular}
\caption{EWs and functional forms of EW ratios in optical and NIR CARMENES spectra that are used to establish best fitting calibration relationships are plotted against the measured $T_{\rm eff}$, radius and luminosity (log L/$L_{\sun}$) for our calibration sample. The data points are color-coded according to the [Fe/H] values taken from Schweitzer et al. (\citet{Schweitzer2019}) and the dashed straight lines represent the linear correlation.}
\label{fig:fig2_EW_ratio_plot}
\end{figure*}

\begin{table*}
\centering
\vspace*{-0.4 cm}
 \caption{Measured EWs of the spectral features from the optical and NIR CARMENES spectra for the interferometric calibration sample.}
  \label{tab:table2_EWs_CalibrationSample}
  \scriptsize
  \hspace*{-0.6 cm}
 \begin{tabular}{lccccccccccc}
  \hline \hline

Star  & Ti (0.844 $\mu$m) & Ca (0.850 $\mu$m) & Ca (0.854 $\mu$m) & Ca (0.866 $\mu$m) & Mg (0.881 $\mu$m) & Ca (1.034 $\mu$m) & Na (1.140 $\mu$m) & K (1.169 $\mu$m) & K (1.243 $\mu$m) & K (1.252 $\mu$m) & Mg (1.574 $\mu$m) \\
              & (\AA) & (\AA) & (\AA) & (\AA) & (\AA) & (\AA) & (\AA) & (\AA) & (\AA) & (\AA) & (\AA) \\
\hline

Gl 172  &	1.231$\pm$0.105 & 0.901$\pm$0.007 & 4.393$\pm$0.310 & 3.530$\pm$0.202 &0.756$\pm$0.030 & 0.797$\pm$0.022 & 2.019$\pm$0.018 &0.620$\pm$0.015 & 0.066$\pm$0.017 & 0.354$\pm$0.011 & 4.971$\pm$0.246 \\ 
Gl 208  &	1.293$\pm$0.055 & 0.803$\pm$0.004 & 4.006$\pm$0.145 & 3.320$\pm$0.107 &0.692$\pm$0.017 &  \textemdash    & 2.340$\pm$0.051 &0.689$\pm$0.008 & 0.188$\pm$0.017 & 0.512$\pm$0.008 & 5.283$\pm$0.127 \\
GJ 809  &	0.573$\pm$0.038 & 0.796$\pm$0.010 & 3.887$\pm$0.187 & 3.028$\pm$0.188 &0.696$\pm$0.025 & 0.817$\pm$0.010 & 3.770$\pm$0.009 &0.986$\pm$0.030 & 0.258$\pm$0.005 & 0.451$\pm$0.016 & 4.644$\pm$0.126 \\
GJ 412A &	1.251$\pm$0.026 & 0.752$\pm$0.007 & 3.565$\pm$0.147 & 2.856$\pm$0.381 &0.534$\pm$0.019 & 0.786$\pm$0.005 & 2.682$\pm$0.021 &0.733$\pm$0.024 & 0.031$\pm$0.013 & 0.302$\pm$0.017 & 2.219$\pm$0.106 \\
GJ 15A  &	1.399$\pm$0.090 & 0.760$\pm$0.026 & 3.253$\pm$0.430 & 2.926$\pm$0.275 &0.463$\pm$0.025 & 0.822$\pm$0.020 & 3.219$\pm$0.018 &0.866$\pm$0.016 & 0.191$\pm$0.021 & 0.375$\pm$0.010 & 2.123$\pm$0.218 \\
GJ 649  &	1.283$\pm$0.029 & 0.794$\pm$0.009 & 3.399$\pm$0.263 & 2.753$\pm$0.211 &0.335$\pm$0.011 & 0.887$\pm$0.017 & 3.646$\pm$0.014 &0.947$\pm$0.025 & 0.245$\pm$0.017 & 0.459$\pm$0.006 & 3.523$\pm$0.035 \\
GJ 205  &	1.439$\pm$0.052 & 0.867$\pm$0.005 & 3.656$\pm$0.174 & 3.209$\pm$0.121 &0.631$\pm$0.016 &  \textemdash    & 4.125$\pm$0.027 &1.120$\pm$0.027 & 0.310$\pm$0.012 & 0.546$\pm$0.015 & 4.829$\pm$0.194 \\
GJ 880  & \textemdash     & 0.641$\pm$0.010 & 3.153$\pm$0.115 & 2.672$\pm$0.273 &0.143$\pm$0.026 & 0.807$\pm$0.021 & 4.489$\pm$0.019 &1.131$\pm$0.028 & 0.288$\pm$0.007 & 0.472$\pm$0.010 & 4.547$\pm$0.141 \\
GJ 526  & 1.075$\pm$0.042 & 0.761$\pm$0.009 & 3.438$\pm$0.266 & 2.114$\pm$0.280 &0.616$\pm$0.015 & 0.773$\pm$0.024 & 3.109$\pm$0.038 &0.776$\pm$0.025 & 0.233$\pm$0.008 & 0.396$\pm$0.005 & 3.524$\pm$0.090 \\
GJ 411  & 0.007$\pm$0.017 & 0.151$\pm$0.007 & 2.616$\pm$0.062 & 2.040$\pm$0.144 &0.064$\pm$0.013 &  \textemdash    & 3.507$\pm$0.010 &0.502$\pm$0.023 & 0.024$\pm$0.008 & 0.063$\pm$0.012 & 2.775$\pm$0.027 \\
GJ 176  & 1.315$\pm$0.071 & 0.679$\pm$0.009 & 3.325$\pm$0.240 & 2.757$\pm$0.141 &0.617$\pm$0.022 & 0.928$\pm$0.028 & 2.152$\pm$0.027 &0.641$\pm$0.029 & 0.011$\pm$0.013 & 0.372$\pm$0.010 & 2.740$\pm$0.261 \\
Gl 104  & 1.247$\pm$0.180 & 0.777$\pm$0.020 & 3.621$\pm$0.455 & 3.036$\pm$0.314 &0.465$\pm$0.049 & 0.900$\pm$0.012 & 2.873$\pm$0.025 &0.577$\pm$0.045 & 0.268$\pm$0.018 & 0.528$\pm$0.014 & 3.113$\pm$0.099 \\
GJ 436  & 1.006$\pm$0.075 & 0.661$\pm$0.012 & 2.593$\pm$0.197 & 2.072$\pm$0.578 &0.500$\pm$0.025 & 0.971$\pm$0.012 & 3.220$\pm$0.011 &0.801$\pm$0.012 & 0.269$\pm$0.023 & 0.476$\pm$0.009 & 2.104$\pm$0.079 \\
GJ 581  & 1.127$\pm$0.062 & 0.560$\pm$0.010 & 2.424$\pm$0.461 & 2.162$\pm$0.308 &0.587$\pm$0.017 & 0.996$\pm$0.010 & 3.909$\pm$0.016 &1.208$\pm$0.068 & 0.318$\pm$0.014 & 0.539$\pm$0.042 & 1.722$\pm$0.293 \\
Gl 109  & 1.389$\pm$0.075 & 0.658$\pm$0.015 & 2.495$\pm$0.221 & 2.445$\pm$0.224 &0.449$\pm$0.037 & 0.983$\pm$0.011 & 4.051$\pm$0.023 &1.105$\pm$0.019 & 0.276$\pm$0.014 & 0.531$\pm$0.008 & 1.579$\pm$0.093 \\
GJ 628  &  \textemdash    & 0.390$\pm$0.011 & 2.548$\pm$0.303 & 1.753$\pm$0.182 &0.135$\pm$0.010 & 0.708$\pm$0.031 & 3.457$\pm$0.029 &0.658$\pm$0.069 & 0.281$\pm$0.026 & 0.530$\pm$0.018 & 1.528$\pm$0.212 \\
GJ 273  & 1.450$\pm$0.065 & 0.531$\pm$0.008 & 2.201$\pm$0.263 & 1.871$\pm$0.156 &0.400$\pm$0.020 & 0.011$\pm$0.010 & 1.543$\pm$0.013 &0.124$\pm$0.009 &   \textemdash   & 0.056$\pm$0.014 & 0.666$\pm$0.044 \\
GJ 699  &   \textemdash   &  \textemdash    & 0.160$\pm$0.290 & 0.642$\pm$0.159 & \textemdash    & 0.260$\pm$0.018 & 2.087$\pm$0.018 &0.025$\pm$0.034 &   \textemdash   & 0.014$\pm$0.021 & 1.473$\pm$0.085 \\
GJ 729  &	1.652$\pm$0.041 & 0.493$\pm$0.012 & 1.898$\pm$0.228 & 2.323$\pm$0.200 &0.467$\pm$0.028 & 1.095$\pm$0.042 & 4.497$\pm$0.021 &1.476$\pm$0.015 & 0.498$\pm$0.016 & 0.824$\pm$0.015 & 1.225$\pm$0.281 \\
GJ 447  & 1.087$\pm$0.101 & 0.700$\pm$0.010 & 3.043$\pm$0.252 & 2.786$\pm$0.734 &0.621$\pm$0.033 & 0.880$\pm$0.025 & 2.466$\pm$0.020 &0.588$\pm$0.019 & 0.249$\pm$0.009 & 0.463$\pm$0.020 & 2.907$\pm$0.346 \\
GJ 876  & 1.410$\pm$0.067 & 0.477$\pm$0.022 & 1.815$\pm$0.251 & 1.847$\pm$0.532 &0.444$\pm$0.043 & 0.980$\pm$0.031 & 4.513$\pm$0.035 &1.182$\pm$0.053 & 0.418$\pm$0.016 & 0.732$\pm$0.017 & 1.469$\pm$0.193 \\
Gl 1253 &  \textemdash    & 0.066$\pm$0.028 & 1.511$\pm$0.781 & 1.421$\pm$0.698 &0.168$\pm$0.110 & 0.627$\pm$0.081 & 3.550$\pm$0.025 &0.409$\pm$0.020 & 0.259$\pm$0.017 & 0.847$\pm$0.018 & 1.091$\pm$0.263 \\
Gl 905  &  \textemdash    &   \textemdash   & 0.456$\pm$0.298 & 0.679$\pm$0.807 &0.135$\pm$0.068 & 0.594$\pm$0.056 & 4.048$\pm$0.051 &0.251$\pm$0.083 &   \textemdash   & 0.366$\pm$0.021 & 0.879$\pm$0.103 \\

  \hline
\end{tabular}
 \vspace{1ex} 
 
 \textbf{Notes.}
     \raggedright Wavelengths are given in vacuum and the respective feature windows and continuums are given in Table \ref{fig:fig1_spectral_features} . \\
     
\end{table*}

\hspace{-0.76 cm} Here,  $F(\lambda)$ represents the flux across the wavelength range of the line $(\lambda_{2} - \lambda_{1} )$, and $F_{c}(\lambda)$ stands for the estimated continuum flux on either side of the absorption line. The EWs are estimated using the modified versions of the publicly available IDL-based \texttt{tellrv}\footnote{\url{https://github.com/ernewton/tellrv}} and \texttt{nirew}\footnote{\url{https://github.com/ernewton/nirew}} packages originally developed by Newton et al. (\citet{Newton2014}, \citet{Newton2015}). M dwarf's spectra are typically polluted by broad absorption features or affected by extinction and contributions from emission sources such as hot dust. So based on our observation, we selected a pseudo continuum for each atomic feature by linear interpolation between adjacent regions, roughly  0.001-0.0015 $\mu$m wide on each side of the absorption line. The features and the continuum points used to calculate the equivalent widths are given in Table \ref{tab:table1_spectralfeature}. The CARMENES high-resolution spectra show all three components of the NIR Mg I triplet (1.5740 $\mu$m, 1.5748 $\mu$m, 1.5765 $\mu$m), and as our continuum window covers all the components, they jointly contribute to the EW estimation of the Mg I (1.57 $\mu$m) feature. We have estimated the uncertainties on the EWs using a Monte Carlo simulation where multiple random realizations of Gaussian noise are added to the spectrum and EWs of the features are recalculated (Khata et al. \citet{Khata2020}). The estimated EWs along with the associated errors of the spectral features from the optical and NIR CARMENES spectra for our interferometric calibration sample are given in Table \ref{tab:table2_EWs_CalibrationSample}.

\subsection{Spectral features selection: Behavior with stellar parameters}

\begin{table}
 \centering
 \caption{Linear Pearson Correlation Coefficient (LPCC) values between the features and stellar parameters for the calibration sample.}
 \hspace{0.0 cm}
 \begin{tabular}{lcccccc}
  \hline \hline
     & & & & LPCC  & & \\
  Feature & & $T_{\rm eff}$ (K) & & Radius ($R_{\sun}$) & & log (L/$L_{\sun}$) \\
  \hline
Ti I (0.844 $\mu$m)  & & -0.17  & & -0.12 & & -0.17 \\
Ca II (0.850 $\mu$m) & & +0.70  & & +0.67 & & +0.69 \\
Ca II (0.854 $\mu$m) & & +0.87  & & +0.83 & & +0.86 \\  
Ca II (0.866 $\mu$m) & & +0.81  & & +0.75 & & +0.78 \\ 
Mg I (0.881 $\mu$m)  & & +0.55  & & +0.43 & & +0.46 \\
Ca I (1.034 $\mu$m)  & & +0.28  & & +0.29 & & +0.31 \\  
Na I (1.140 $\mu$m)  & & -0.23  & & -0.06 & & -0.10 \\
K I (1.169 $\mu$m)   & & +0.26  & & +0.30 & & +0.32 \\
K I (1.243 $\mu$m)   & & -0.49  & & -0.32 & & -0.39 \\
K I (1.252 $\mu$m)   & & -0.15  & & -0.04 & & -0.09 \\
Mg I (1.574 $\mu$m)  & & +0.88  & & +0.85 & & +0.82 \\

   \hline
 \end{tabular}
 \vspace{1ex} \\
   \raggedright \texttt{{\scriptsize {Notes : Wavelengths are given in vacuum}}}.\\
    \label{tab:table3_lpcc}
\end{table}

We performed a principal component analysis (PCA) to search for the strongest correlations between the EWs of our calibration sample, and we calculated the linear Pearson correlation coefficients (LPCC) looking for legitimate correlation with the EWs and $T_{\rm eff}$, radius, and luminosity. The LPCC value ranges from -1 to +1, and for two variables, such as X and Y, a positive correlation coefficient value indicates Y is increasing as X increases, and a negative value means Y decreases as X increases. We found that the EWs of Ca II (0.854 $\mu$m) and Ca II (0.866 $\mu$m) in the optical channel and the EWs of Mg I (1.574 $\mu$m) in the NIR channel attribute mostly in the first principal components (explaining 79\% of the variance) in the PCA test. These three features also give the maximum LPCC values while we investigated for linear correlation with the stellar parameters, and we have given the respective LPCC values of the spectral features in Table \ref{tab:table3_lpcc}. These two singly ionized Ca II (0.854 $\mu$m, 0.866 $\mu$m) lines (e.g., Ca II IRT infrared triplet) are prominent features in the spectra of cool M type stars and lie in a region of the spectrum with a well-defined continuum and are significantly free from telluric lines. Earlier work has helped to establish the sensitivity of these lines with stellar atmospheric parameters, such as Mallik  (\citet{Mallik1994}, \citet{Mallik1997}) and Andretta et al. (\citet{Andretta2005}) tried to model the behavior of Ca II IRT line strength with $T_{\rm eff}$, log g and metallicity; Zhou  (\citet{Zhou1991}) showed a significant effect of temperatures on Ca II IRT lines for cooler stars (up to M7) at low $T_{\rm eff}$ and Terrien et al. \citet{Terrien2015} developed a technique to determine M-dwarfs M$_{k}$ and radius using Ca II IRT feature. Among all the features in the optical and NIR band, Mg I (1.574 $\mu$m) shows the best correlation with the stellar parameters. The EW of the Mg I (1.574 $\mu$m) feature is used by Newton et al. (\citet{Newton2015}) to estimate the stellar radius of M dwarfs in the MEarth sample; Martinez et al. (\citet{Martinez2017}) used this feature to calibrate $T_{\rm eff}$ and radius of K- and M-dwarf stars; and later the feature is also used by Khata et al. (\citet{Khata2020}) for the calibration of $T_{\rm eff}$, radius and luminosity of low-resolution M-dwarfs spectra. Therefore, based on the PCA test and the best LPCC values obtained, we selected these three spectral features from the initially considered 11 features for empirical calibrations of the stellar parameters.  \\
$~~~~~$ In Fig. \ref{fig:fig2_EW_ratio_plot}, we show the plot of the EWs and ratios of EWs for the selected spectral features [i.e., Ca (0.854 $\mu$m), Ca (0.866 $\mu$m), Mg (1.574 $\mu$m)] for the interferometric calibration sample against $T_{\rm eff}$, radius and luminosity. The data points are color-coded by their respective metallicity ([Fe/H]) values as estimated by Schweitzer et al. (\citet{Schweitzer2019}). As the atomic lines get broadened from early to later M dwarfs, the EWs decrease as effective temperature decreases. We see that the EWs of the two Ca II lines and the Mg I line and the line ratios behave almost in a linear fashion, implied by the dashed straight line fittings, as functions of $T_{\rm eff}$, radius and log (L/$L_{\sun}$) with no statistically significant metallicity dependence.

\begin{table*}
\centering
\vspace*{-0.4 cm}
\caption{Different types of linear trial functions with one example each, using simple functional forms of the selected EWs and EW ratios (x = EW$_{Ca [0.854 \mu m]}$ ; y = EW$_{Ca [0.866 \mu m]}$ ; z = EW$_{Mg [1.574 \mu m]}$) and associated RMSE, MAD and ${R^2}_{ap}$ values of the stellar parameters. The upper case X, Y, Z represent functional forms of x, y, z ( i.e, X = x , $x^2$ , $x^{1/2}$ , 1/x , $1/x^2$ , $1/x^{1/2}$ ; Y = y , $y^2$ , $y^{1/2}$ , 1/y , $1/y^2$ , $1/y^{1/2}$ ; Z = z , $z^2$ , $z^{1/2}$ , 1/z , $1/z^2$ , $1/z^{1/2}$).}
\label{tab:table4_trial_fun}
\hspace*{-1.05 cm}
 \begin{tabular}{lccccccccccc}
  \hline \hline

& &&   $T_{\rm eff}$ (K) &&&  \hspace{-0.3cm} Radius (${R}_\odot$) &&& \hspace{-0.3cm} log (L/${L}_\odot$) &   & \\ 

 Nature of linear test functions  & Example function & RMSE & MAD & ${R^2}_{ap}$  & \hspace{0.3cm} RMSE & \hspace{-0.8cm} MAD & \hspace{-1.0cm} ${R^2}_{ap}$ & \hspace{0.0cm} RMSE & \hspace{-0.8cm} MAD & \hspace{-0.7cm} ${R^2}_{ap}$ & \\
 
\hline
\hline
single EW, single component [a + bX] &  a+bx &147.15&  111.90&   0.737&   \hspace{0.3cm} 0.082&  \hspace{-0.8cm} 0.062&\hspace{-1.0cm}   0.672  &   0.253& \hspace{-0.8cm} 0.174& \hspace{-0.7cm}   0.730& \\
single EW, single component [a + bY] &  a+by &172.40&  140.07&   0.638&  \hspace{0.3cm}  0.096& \hspace{-0.8cm}0.071& \hspace{-1.0cm}  0.547&   0.311&\hspace{-0.8cm}  0.216& \hspace{-0.7cm}   0.591& \\
single EW, single component [a + bZ] &  a+bz & 137.18&  108.68&   0.771& \hspace{0.3cm} 0.076& \hspace{-0.8cm} 0.055& \hspace{-1.0cm}  0.712&   0.282&\hspace{-0.8cm}  0.216& \hspace{-0.7cm}   0.663& \\
\hline

single EW, double component [a + bX +c$X^{\prime}$ ($X^{\prime}\subseteq$X; $X^{\prime}\neq$X)] &a+bx+c$x^{2}$       &126.75&  98.00&   0.805&   \hspace{0.3cm}    0.078&\hspace{-0.8cm}  0.053& \hspace{-1.0cm}  0.701&        0.249& \hspace{-0.8cm} 0.171& \hspace{-0.7cm}   0.737& \\
single EW, double component [a + bY +c$Y^{\prime}$ ($Y^{\prime}\subseteq$Y; $Y^{\prime}\neq$Y)] &a+by+c$y^{2}$       &166.72&  132.76&   0.662&  \hspace{0.3cm}    0.097& \hspace{-0.8cm} 0.066& \hspace{-1.0cm}  0.540&     0.318& \hspace{-0.8cm} 0.215&  \hspace{-0.7cm}  0.573& \\
single EW, double component [a + bZ +c$Z^{\prime}$ ($Z^{\prime}\subseteq$Z; $Z^{\prime}\neq$Z)] &a+bz+c$z^{2}$       &138.66&  109.38&   0.766&   \hspace{0.3cm}   0.078& \hspace{-0.8cm} 0.053& \hspace{-1.0cm}  0.699&     0.283& \hspace{-0.8cm} 0.195& \hspace{-0.7cm}   0.661& \\
 \hline      
    
double EWs, double component [a +bX +cY] &a+bx+cy     &150.66  &112.24  &0.724        & \hspace{0.3cm}  0.083  &\hspace{-0.8cm}0.062  &\hspace{-1.0cm}0.663         & 0.254  &\hspace{-0.8cm} 0.175  &\hspace{-0.7cm} 0.726& \\
double EWs, double component [a +bX +cZ] & a+bx+cz  &114.20  &85.69   &0.841        &  \hspace{0.3cm} 0.069  &\hspace{-0.8cm} 0.041  &\hspace{-1.0cm} 0.769         & 0.230  &\hspace{-0.8cm} 0.139  &\hspace{-0.7cm} 0.777& \\
double EWs, double component [a +bY +cZ]  &a+by+cz &124.14  &98.57   &0.812         &\hspace{0.3cm}  0.074  & \hspace{-0.8cm} 0.046  & \hspace{-1.0cm} 0.730         & 0.261  &\hspace{-0.8cm} 0.172  &\hspace{-0.7cm} 0.712& \\
 \hline      

single EW ratio, single component [a +b$R_{1}$ ($R_{1}$ =X/Y)] &a+bx/y          & 248.36  &191.47 &  0.250           &\hspace{0.3cm} 0.119 & \hspace{-0.8cm} 0.095& \hspace{-1.0cm} 0.299       & 0.390& \hspace{-0.8cm} 0.309& \hspace{-0.7cm}   0.357&  \\
single EW ratio, single component [a +b$R_{2}$ ($R_{2}$ =X/Z)] &a+bx/z & 285.48  &209.99 &  0.008          & \hspace{0.3cm} 0.143 &\hspace{-0.8cm} 0.112&\hspace{-1.0cm} -0.008       & 0.493& \hspace{-0.8cm} 0.383& \hspace{-0.7cm}  -0.029&  \\
single EW ratio, single component [a +b$R_{3}$ ($R_{3}$ =Y/Z)] &a+by/z & 264.92  &193.59 &  0.146          &\hspace{0.3cm} 0.134 & \hspace{-0.8cm} 0.106& \hspace{-1.0cm} 0.115       & 0.466& \hspace{-0.8cm} 0.362& \hspace{-0.7cm}   0.081&  \\
 \hline      

single EW $\&$ single EW ratio, double component [a+bX+c$R_{1}$] & a+bx+cx/y &140.03 & 110.91  & 0.761   &\hspace{0.3cm} 0.083 &\hspace{-0.8cm} 0.060 & \hspace{-1.0cm} 0.662     & 0.258 & \hspace{-0.8cm} 0.171 & \hspace{-0.7cm} 0.719  &  \\
single EW $\&$ single EW ratio, double component [a+bX+c$R_{2}$]  & a+bx+cx/z &128.86 & 100.45  & 0.798   &\hspace{0.3cm} 0.077 & \hspace{-0.8cm}0.052 &\hspace{-1.0cm}  0.710     & 0.246 & \hspace{-0.8cm} 0.159 & \hspace{-0.7cm} 0.745  &  \\
single EW $\&$ single EW ratio, double component [a+bX+c$R_{3}$]  & a+bx+cy/z &128.28 & 101.01  & 0.800   &\hspace{0.3cm} 0.076 & \hspace{-0.8cm} 0.053 &\hspace{-1.0cm}  0.715     & 0.241 & \hspace{-0.8cm} 0.156 & \hspace{-0.7cm} 0.753  &  \\
 \hline      

single EW $\&$ single EW ratio, double component [a+bY+c$R_{1}$]  &a+by+cx/y  &173.69 & 133.40 &  0.633   &\hspace{0.3cm} 0.094 & \hspace{-0.8cm} 0.072 &\hspace{-1.0cm}  0.566     & 0.296 &\hspace{-0.8cm} 0.208 & \hspace{-0.7cm} 0.629  &  \\
single EW $\&$ single EW ratio, double component [a+bY+c$R_{2}$]  &a+by+cx/z &159.53 & 124.71 &  0.690   & \hspace{0.3cm} 0.093 & \hspace{-0.8cm} 0.064 & \hspace{-1.0cm} 0.574     & 0.309 &\hspace{-0.8cm} 0.204 & \hspace{-0.7cm} 0.596  &  \\
single EW $\&$ single EW ratio, double component [a+bY+c$R_{3}$]  &a+by+cy/z  &143.75 & 109.19 &  0.749   & \hspace{0.3cm} 0.087 & \hspace{-0.8cm} 0.057 & \hspace{-1.0cm} 0.631     & 0.288 &\hspace{-0.8cm} 0.185 & \hspace{-0.7cm} 0.648  &  \\
 \hline

single EW $\&$ single EW ratio, double component [a+bZ+c$R_{1}$]  &a+bz+cx/y  &130.58  &101.53 &  0.793   &\hspace{0.3cm} 0.070 & \hspace{-0.8cm} 0.048 & \hspace{-1.0cm} 0.758     & 0.245 &\hspace{-0.8cm} 0.174 & \hspace{-0.7cm} 0.745  &  \\
single EW $\&$ single EW ratio, double component [a+bZ+c$R_{2}$] &a+bz+cx/z  &124.47  &98.52  &  0.811   &\hspace{0.3cm} 0.070 & \hspace{-0.8cm} 0.044 & \hspace{-1.0cm} 0.762     & 0.246 &\hspace{-0.8cm} 0.172 & \hspace{-0.7cm} 0.744  &  \\
single EW $\&$ single EW ratio, double component [a+bZ+c$R_{3}$]  &a+bz+cy/z &132.00  &105.23 &  0.788   &\hspace{0.3cm} 0.074 & \hspace{-0.8cm} 0.049 & \hspace{-1.0cm} 0.733     & 0.268 &\hspace{-0.8cm} 0.197 & \hspace{-0.7cm} 0.697  &  \\
 \hline      

single EW ratio, double component [a +bR$_{1}$ +cR$_{1}^{\prime}$ (R$_{1}^{\prime}\neq R_{1}$)] &  a+bx/y+cx/$y^{2}$	  &181.69  &138.02   &0.598      &\hspace{0.3cm} 0.100  &\hspace{-0.8cm} 0.078  & \hspace{-1.0cm} 0.507    & 0.305  &\hspace{-0.8cm} 0.222   &\hspace{-0.7cm} 0.605  &  \\ 
single EW ratio, double component [a +bR$_{2}$ +cR$_{2}^{\prime}$ (R$_{2}^{\prime}\neq R_{2}$)] & a+bx/z+cx/$z^{2}$	&261.95  &196.47  & 0.165 &\hspace{0.3cm} 0.136 & \hspace{-0.8cm} 0.111 & \hspace{-1.0cm}  0.082    & 0.453 &\hspace{-0.8cm} 0.362 &\hspace{-0.7cm}  0.131  &  \\
single EW ratio, double component [a +bR$_{3}$ +cR$_{3}^{\prime}$ (R$_{3}^{\prime}\neq R_{3}$)] & a+by/z+cy/$z^{2}$	&262.06  &189.85  & 0.164    &\hspace{0.3cm} 0.136 & \hspace{-0.8cm} 0.104  & \hspace{-1.0cm} 0.093   & 0.464  &\hspace{-0.8cm} 0.352  & \hspace{-0.7cm} 0.088  &  \\
 \hline 
   
double EW ratio, double component [a + bR$_{1}$ + cR$_{2}$] & a+bx/y+cx/z &222.67  &168.38  &  0.397    &\hspace{0.3cm} 0.108  &\hspace{-0.8cm} 0.085  &\hspace{-1.0cm} 0.426    & 0.361  &\hspace{-0.8cm} 0.282  & \hspace{-0.7cm} 0.447  &  \\
double EW ratio, double component [a + bR$_{1}$ + cR$_{3}$] & a+bx/y+cy/z  &220.25  &165.42  &  0.410     &\hspace{0.3cm} 0.108  &\hspace{-0.8cm} 0.084  &\hspace{-1.0cm}  0.429    & 0.359  &\hspace{-0.8cm} 0.274  &\hspace{-0.7cm}  0.455  &  \\    
double EW ratio, double component [a + bR$_{2}$ + cR$_{3}$] &a+bx/z+cy/z &229.74  &171.63 &  0.358  &\hspace{0.3cm} 0.116  &\hspace{-0.8cm} 0.093 & \hspace{-1.0cm} 0.335  & 0.385  &\hspace{-0.8cm} 0.299 & \hspace{-0.7cm} 0.372  &  \\ 
 
\hline
\hline
 \multicolumn{10}{c}{Accepted linear functions for $T_{eff}$ (${\star}$), Radius (${\dagger}$) and Luminosity (${\mathsection}$) respectively with minimum RMSE and maxmimum ${R^2}_{ap}$ values} \\
 \hline
 \hline
 
(${\star}$) single EW $\&$ single EW ratio, double component  &   a+bx+c$z^{2}$/x & 99.00  & 70.97   & 0.881   &\hspace{0.3cm}  & \hspace{-0.8cm}  & \hspace{-1.0cm}     &  & \hspace{-0.8cm}  & \hspace{-0.7cm}   &  \\ 
(${\dagger}$) single EW $\&$ single EW ratio, double component &   a+bx+c$z^{2}$/y 	   		& &    &    & \hspace{0.3cm} 0.067 &\hspace{-0.8cm} 0.043 & \hspace{-1.0cm} 0.779    &  &\hspace{-0.8cm}  & \hspace{-0.7cm}   &  \\
(${\mathsection}$) double EW ratio, double component &  a+bz/$\sqrt{y}$+cx/$\sqrt{z}$   &  & &    &\hspace{0.3cm}   &\hspace{-0.8cm}  & \hspace{-1.0cm}    & 0.218  &\hspace{-0.8cm} 0.137 & \hspace{-0.7cm} 0.798  &  \\
\hline
\end{tabular}
 \vspace{1ex} 
 
 \textbf{Notes.}
     \raggedright Table \ref{tab:table4_trial_fun} is available in its entirety in the electronic version of the journal as supplementary material. \\
     
\end{table*}

\begin{figure*}
\centering
\begin{tabular}{c}
\hspace*{-1.0 cm} \includegraphics[width= 7.5 in, height = 4.05 in, clip]{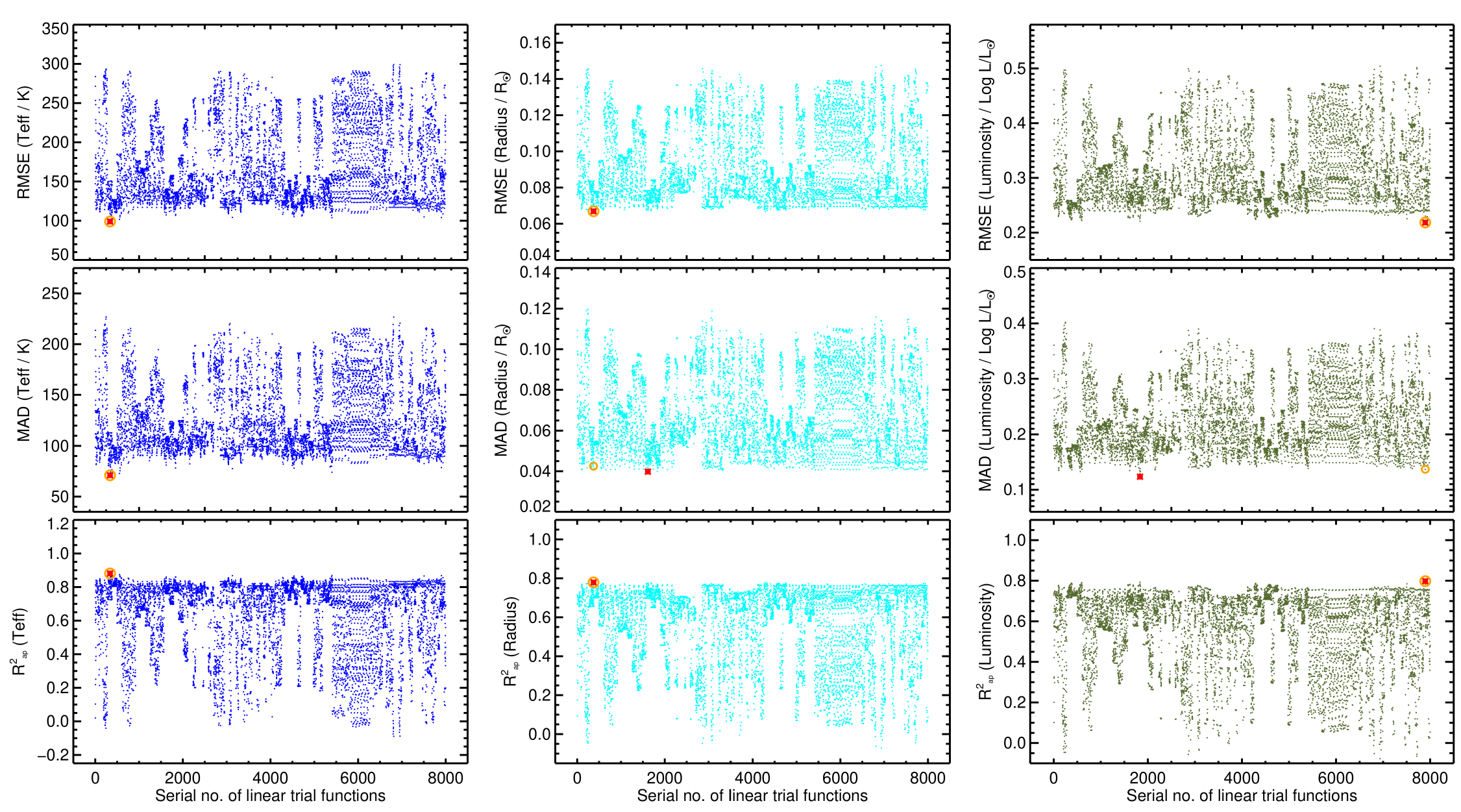}
\end{tabular}
\caption{A representation of the estimated RMSE, MAD and $R^{2}_{ap}$ (plotted along the y-axes) of the different trial functions (plotted along the x-axes going from 1 to 8001) for $T_{\rm eff}$ (left panel, blue color), radius (middle panel, cyan color) and log luminosity (right panel, olive color) is shown here. The respective RMSE, MAD and $R^{2}_{ap}$ values of the chosen best fitting functions are marked by orange circles and the lowest values of RMSE and MAD and the highest value of $R^{2}_{ap}$ are represented by the red star symbols.}
\label{fig:fig3_trial_fun}
\end{figure*}

\subsection{Empirical Calibrations of $T_{\rm eff}$, radius and luminosity}

\begin{figure*}
\centering
\vspace*{-0.3 cm}
\begin{tabular}{ccc}
\includegraphics[width= 3.0 in, height = 2.5 in, clip]{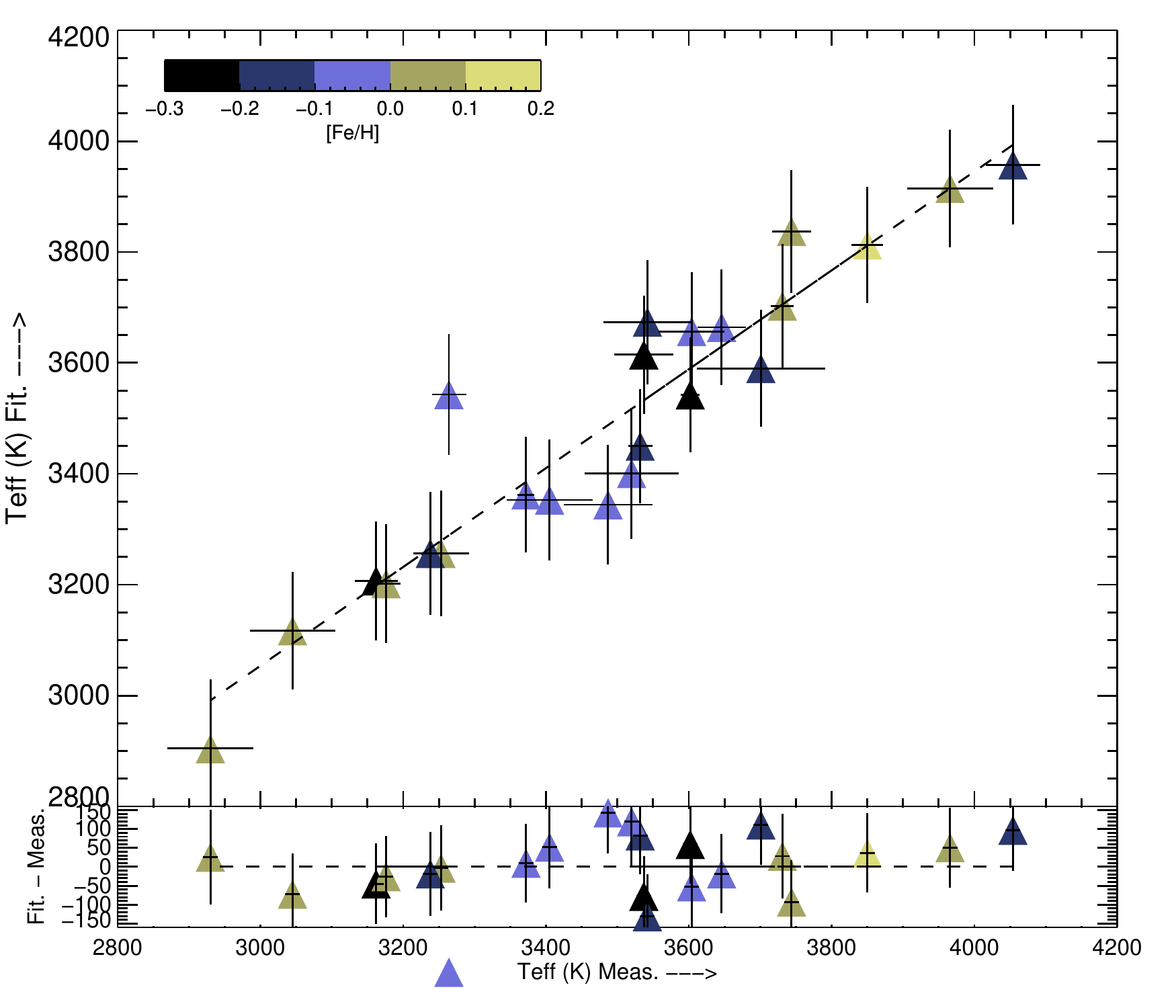} \hspace*{0.7 cm}
\includegraphics[width= 3.0 in, height = 2.5 in, clip]{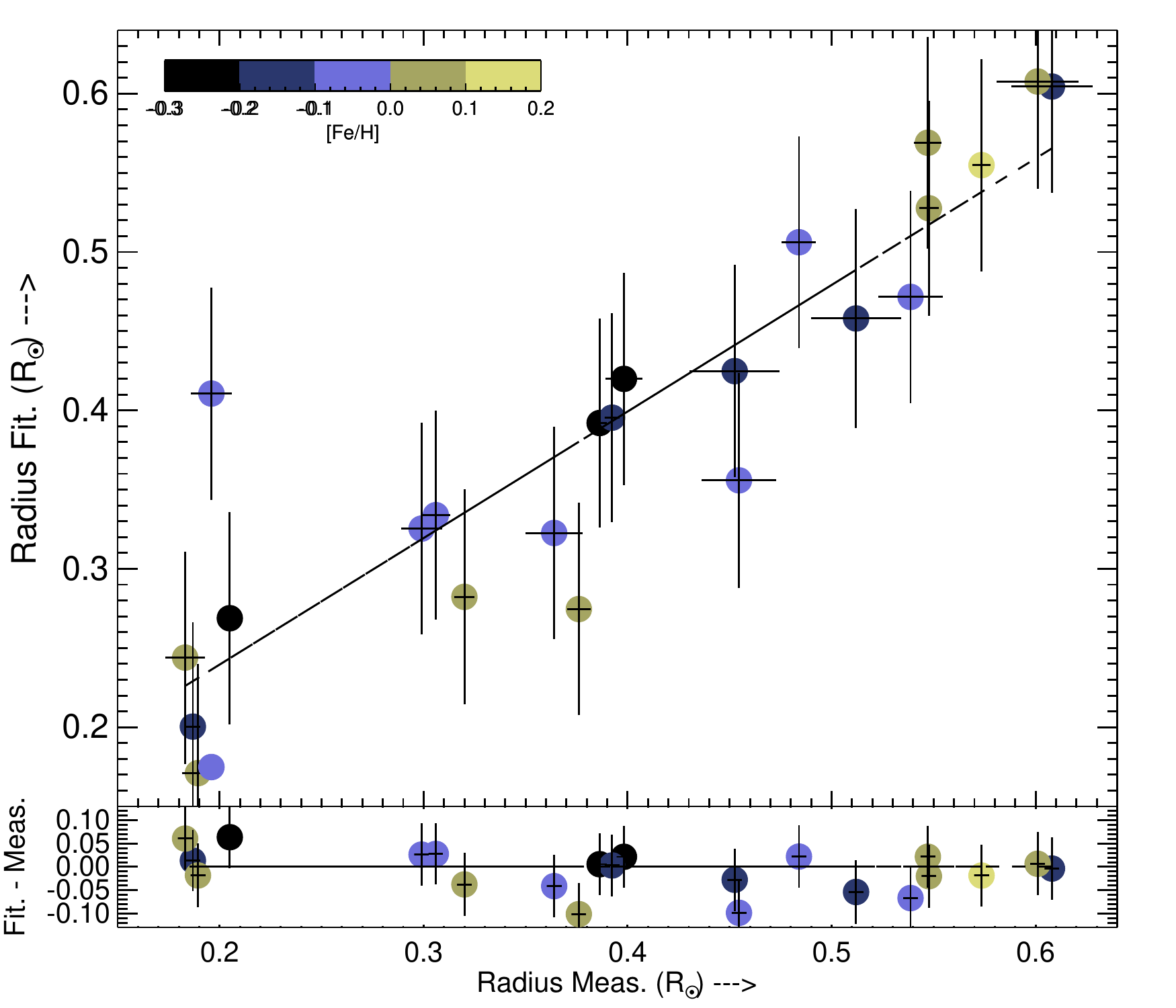}\\[2\tabcolsep]
\includegraphics[width= 3.0 in, height = 2.5 in, clip]{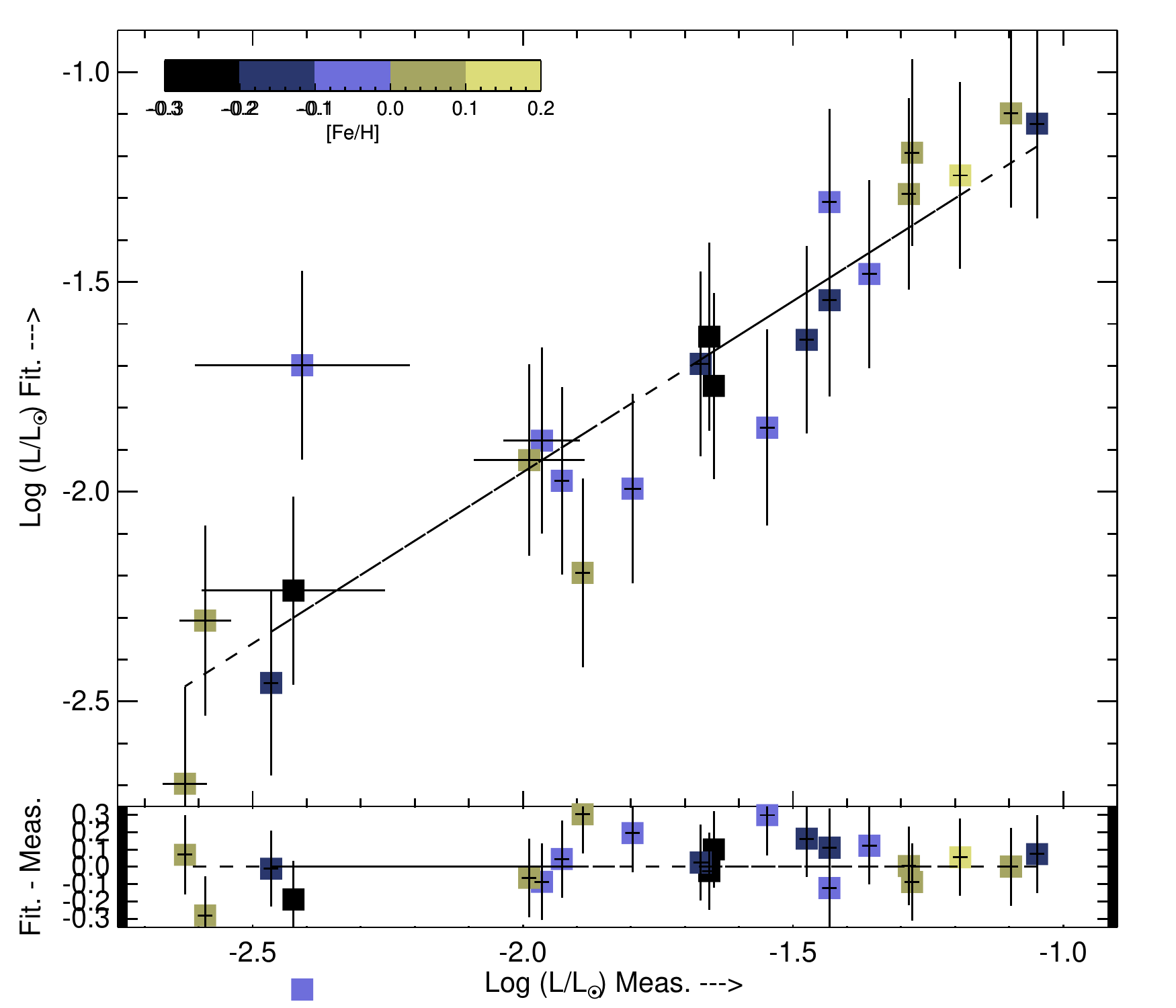}
\end{tabular}
\caption{We show the plots of our best-fitting calibration relationships and the respective residuals for $T_{\rm eff}$ (filled triangles in the top left panel), radius (filled circles in the top right panel), and Log ($L_{bol}$) (filled squares in bottom panel) here. The horizontal axes show the directly measured stellar parameters taken from Boyajian et al. (\citet{Boyajian2012}), Mann et al. (\citet{Mann2013b}, \citet{Mann2015}) and Markus Rabus et al. (\citet{Markus2019}). In the top plot of each panel, the vertical axes represent the stellar parameters inferred from the best fits of this work; and in the lower plot of each panel, the vertical axes represent the residuals between the best-fitting values and the measured values. The data points are color-coded by the [Fe/H] values from Schweitzer et al. (\citet{Schweitzer2019}).}
\label{fig:fig4_parameter_calibration}
\end{figure*}

\begin{table*}
\vspace*{-0.4 cm}
 \centering
 \caption{Inferred $T_{\rm eff}$, radius and $L_{bol}$ for the calibration sample.}
  \label{tab:table5_teff_rad_lum_cal}
 \begin{tabular}{lccccccccc}
  \hline \hline
\large{Star} &  &&\hspace*{-1.95cm} \large{Teff (K)} &  &&\hspace*{-1.95cm} \large{Radius ($R_{\sun}$)}  &  && \hspace*{-1.95cm} \large{log $L/L_{\sun}$}   \\
   && Measured & Inferred $^{(e)}$ && Measured & Inferred $^{(f)}$ && Measured & Inferred $^{(g)}$  \\

 \hline

Gl 172 $^{a}$ && 4054$\pm$38 & 3957$\pm$108 && 0.608$\pm$0.020 &0.604$\pm$0.067 &&\textendash 1.048$\pm$0.009 &\textendash 1.12$\pm$0.23 \\ 
Gl 208 $^{a}$ && 3966$\pm$60 & 3915$\pm$106 && 0.601$\pm$0.020 &0.608$\pm$0.068 &&\textendash 1.097$\pm$0.010 &\textendash 1.10$\pm$0.23 \\
GJ 809 $^{b}$ && 3744$\pm$27 & 3837$\pm$111 && 0.547$\pm$0.006 &0.569$\pm$0.067 &&\textendash 1.280$\pm$0.005 &\textendash 1.19$\pm$0.22 \\
GJ 412A $^{b}$ && 3537$\pm$41 & 3615$\pm$107 && 0.398$\pm$0.009 &0.420$\pm$0.067 &&\textendash 1.655$\pm$0.004 &\textendash 1.63$\pm$0.22 \\
GJ 15A $^{b}$ && 3602$\pm$13 & 3542$\pm$104 && 0.386$\pm$0.002 &0.392$\pm$0.066 &&\textendash 1.647$\pm$0.005 &\textendash 1.75$\pm$0.22 \\
GJ 649 $^{d}$ && 3604$\pm$46 & 3657$\pm$107 && 0.539$\pm$0.015 &0.472$\pm$0.067 &&\textendash 1.359$\pm$0.002 &\textendash 1.48$\pm$0.22 \\
GJ 205 $^{b}$ && 3850$\pm$22 & 3813$\pm$105 && 0.574$\pm$0.004 &0.555$\pm$0.067 &&\textendash 1.190$\pm$0.009 &\textendash 1.25$\pm$0.22 \\
GJ 880 $^{b}$ && 3731$\pm$16 & 3703$\pm$112 && 0.548$\pm$0.004 &0.528$\pm$0.068 &&\textendash 1.286$\pm$0.004 &\textendash 1.29$\pm$0.23 \\
GJ 526 $^{b}$ && 3646$\pm$34 & 3664$\pm$105 && 0.484$\pm$0.008 &0.506$\pm$0.067 &&\textendash 1.433$\pm$0.006 &\textendash 1.31$\pm$0.22 \\
GJ 411 $^{b}$ && 3532$\pm$17 & 3450$\pm$103 && 0.392$\pm$0.003 &0.395$\pm$0.066 &&\textendash 1.671$\pm$0.006 &\textendash 1.70$\pm$0.22 \\
GJ 176 $^{d}$ && 3701$\pm$90 & 3590$\pm$106 && 0.453$\pm$0.022 &0.425$\pm$0.067 &&\textendash 1.475$\pm$0.003 &\textendash 1.64$\pm$0.22 \\
Gl 104 $^{a}$ && 3542$\pm$62 & 3674$\pm$112 && 0.512$\pm$0.022 &0.458$\pm$0.069 &&\textendash 1.432$\pm$0.012 &\textendash 1.54$\pm$0.23 \\
GJ 436 $^{b}$ && 3520$\pm$66 & 3401$\pm$118 && 0.455$\pm$0.018 &0.356$\pm$0.068 &&\textendash 1.548$\pm$0.011 &\textendash 1.85$\pm$0.23 \\
GJ 581 $^{b}$ && 3487$\pm$62 & 3344$\pm$108 && 0.299$\pm$0.010 &0.325$\pm$0.067 &&\textendash 1.928$\pm$0.007 &\textendash 1.97$\pm$0.22 \\
Gl 109 $^{a}$ && 3405$\pm$60 & 3353$\pm$110 && 0.364$\pm$0.014 &0.323$\pm$0.067 &&\textendash 1.797$\pm$0.015 &\textendash 1.99$\pm$0.23 \\
GJ 628 $^{c}$ && 3372$\pm$12 & 3362$\pm$104 && 0.306$\pm$0.007 &0.334$\pm$0.066 &&\textendash 1.965$\pm$0.070 &\textendash 1.88$\pm$0.22 \\
GJ 273 $^{c}$ && 3253$\pm$39 & 3256$\pm$113 && 0.320$\pm$0.005 &0.282$\pm$0.068 &&\textendash 1.988$\pm$0.102 &\textendash 1.92$\pm$0.23 \\
GJ 699 $^{b}$ && 3238$\pm$11 & 3256$\pm$111 && 0.187$\pm$0.001 &0.200$\pm$0.066 &&\textendash 2.466$\pm$0.003 &\textendash 2.46$\pm$0.22 \\
GJ 729 $^{c}$ && 3162$\pm$30 & 3207$\pm$107 && 0.205$\pm$0.006 &0.269$\pm$0.067 &&\textendash 2.424$\pm$0.169 &\textendash 2.24$\pm$0.22 \\
GJ 447 $^{c}$ && 3264$\pm$24 & 3543$\pm$109 && 0.196$\pm$0.010 &0.411$\pm$0.067 &&\textendash 2.408$\pm$0.198 &\textendash 1.70$\pm$0.23 \\
GJ 876 $^{d}$ && 3176$\pm$20 & 3202$\pm$107 && 0.376$\pm$0.005 &0.275$\pm$0.067 &&\textendash 1.889$\pm$0.002 &\textendash 2.19$\pm$0.23 \\
GL 1253 $^{a}$ && 3045$\pm$60 & 3117$\pm$107 && 0.183$\pm$0.009 &0.244$\pm$0.067 &&\textendash 2.588$\pm$0.047 &\textendash 2.31$\pm$0.23 \\
Gl 905 $^{a}$ && 2930$\pm$60 & 2904$\pm$125 && 0.189$\pm$0.007 &0.171$\pm$0.069 &&\textendash 2.626$\pm$0.041 &\textendash 2.70$\pm$0.23 \\
  \hline
\end{tabular}
 \vspace{1ex} 
 
 \textbf{Notes.}
     \raggedright \scriptsize Interferometrically measured parameters are taken from $^{a}$ Mann et al. (\citet{Mann2015}) ; $^{b}$ Newton et al. (\citet{Newton2015}) ; $^{c}$ Markus Rabus et al. (\citet{Markus2019}) ; $^{d}$ Mann et al. (\citet{Mann2013b}).  $^{(e)}$ $T_{\rm eff}$ is inferred using equation (\ref{eq2}), the fitting is shown in Fig. \ref{fig:fig4_parameter_calibration}, top left panel ; $^{(f)}$ Radius is inferred using equation (\ref{eq3}), the fitting is shown in Fig. \ref{fig:fig4_parameter_calibration}, top right panel ; $^{(g)}$ log $L/L_{\sun}$ is inferred using equation (\ref{eq4}), the fitting is shown in Fig. \ref{fig:fig4_parameter_calibration}, bottom panel. \\
     
\end{table*}

\begin{figure*}
\centering
\begin{tabular}{c}
\hspace*{-0.5 cm} \includegraphics[width= 7.2 in, height = 2.3 in, clip]{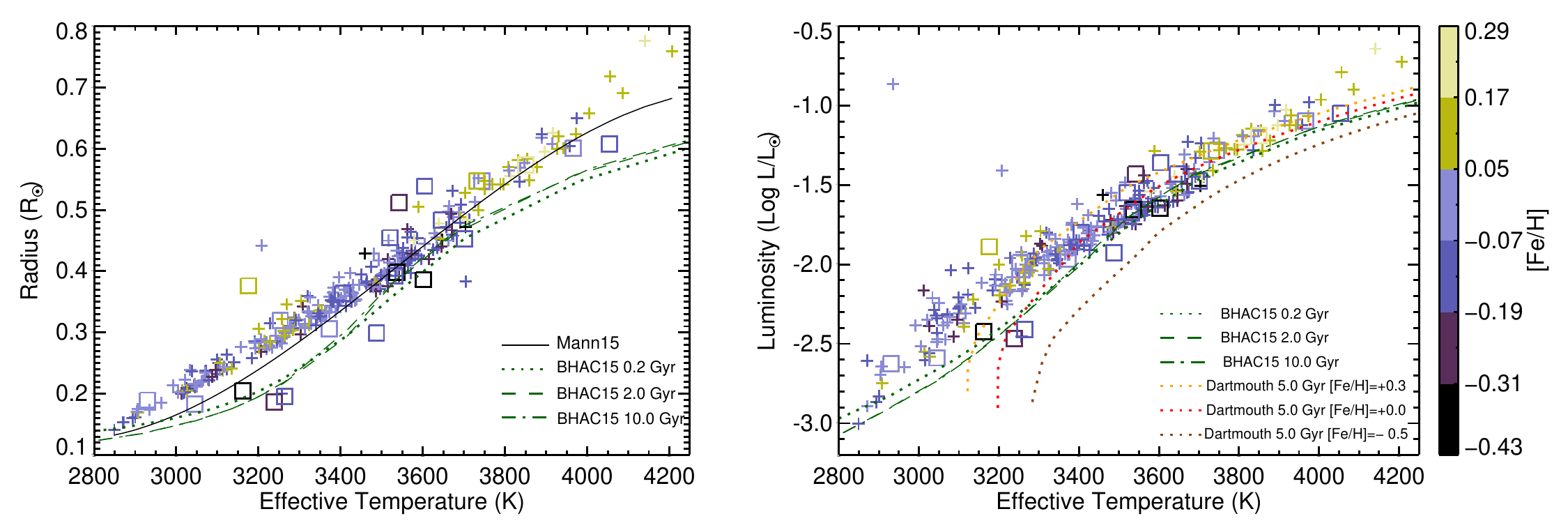}
\end{tabular}
\caption{Stellar radius vs. effective temperature (left panel) and luminosity vs. effective temperature (right panel) plot, showing parameter values of CARMENES M-dwarfs (plus symbols) estimated using EW dependent calibration relations and measured values of interferometric calibrators (open square symbols). For the radius vs $T_{\rm eff}$ plot, we overplot the best-fit radius-temperature relation (solid black line) from Mann15 (Mann et al. \citet{Mann2015}) and 0.2 Gyr, 2.0 Gyr and 10.0 Gyr isochrones (represented by green dotted, dashed and dash-dot lines respectively) for [Fe/H] = 0.0 and log g $\approx$ 5.0 from BHAC15 (Baraffe et al. \citet{Baraffe2015}). For the luminosity-$T_{\rm eff}$ plot, we also show the isochrones from BHAC15 along with the 5 Gyr Dartmouth isochrones for [Fe/H] = 0.3 (dotted orange line), [Fe/H] = 0.0 (dotted red line) and [Fe/H] = - 0.5 (dotted brown line). The data points are color-coded by metallicity from Schweitzer et al. (\citet{Schweitzer2019}).}
\label{fig:fig5_radvsteff_lumvsteff_plt}
\end{figure*}

\begin{figure*}
\centering
\begin{tabular}{c}
\hspace*{-0.6cm} \includegraphics[width= 7.3 in, height = 1.8 in, clip]{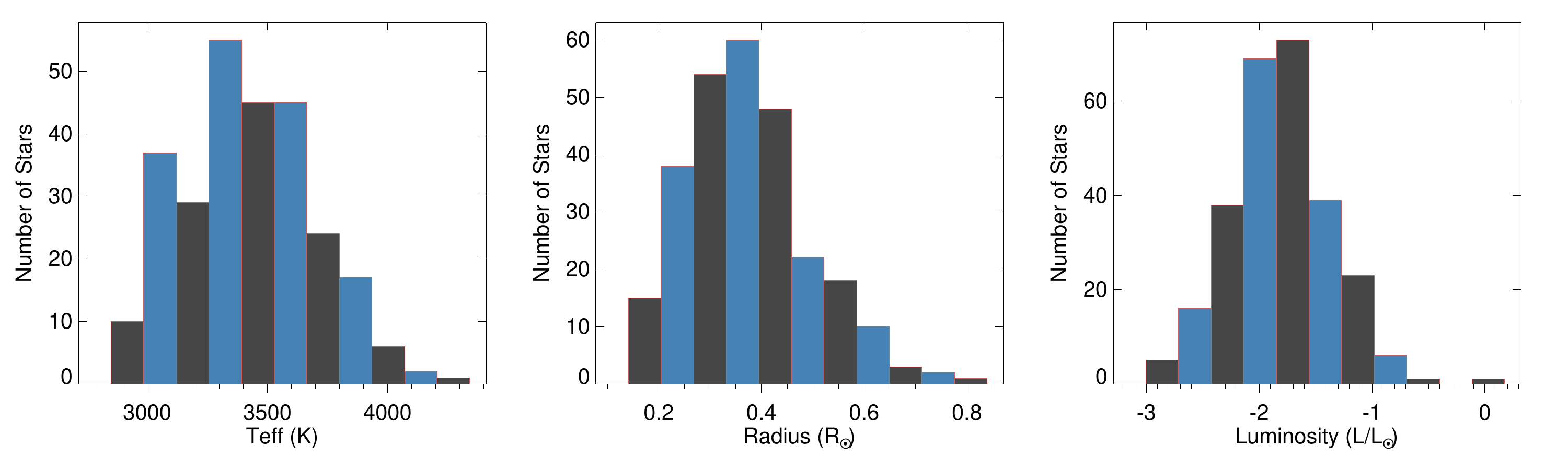}
\end{tabular}
\caption{The distribution of estimated $T_{\rm eff}$ (2849 < $T_{\rm eff}$(K) < 4206 ), radius (0.141 < $R/R_{\sun}$ < 0.776) and luminosity (\textendash 3.00 < log $L/L_{\sun}$ < \textendash 0.643) of the CARMENES M-dwarfs are shown here.}
\label{fig:fig6_para_dist}
\end{figure*}

While looking for the best-fit calibration relations, we found that simple linear functions of EWs or ratios of EW give better results than any higher order polynomial functions (e.g, Hoerl function [a$b^{r}*r^{c}$], power law [a$r^{b}$ ; b > 2], exponential [a$b^{r}$], logarithmic [a + b ln(r)] ; r = EW or ratio of EW). The same kind of linear fits are also observed in the work of Newton et al. (\citet{Newton2015}) but unlike them we did not restrict our fits to only the optical or NIR regions, rather we performed a comprehensive search by combining features from both bands simultaneously. For single-line EWs (we used lower case x, y, z as EWs: x = EW$_{Ca [0.854 \mu m]}$ ; y = EW$_{Ca [0.866 \mu m]}$ ; z = EW$_{Mg [1.574 \mu m]}$), we tested with single component linear functions (e.g., a + bX ; a + bY ; a + bZ), where we used upper case X, Y, Z as simple functional forms of x, y, z ( i.e, X = x , $x^2$ , $x^{1/2}$ , 1/x , $1/x^2$ , $1/x^{1/2}$ ; Y = y , $y^2$ , $y^{1/2}$ , 1/y , $1/y^2$ , $1/y^{1/2}$ ; Z = z , $z^2$ , $z^{1/2}$ , 1/z , $1/z^2$ , $1/z^{1/2}$). We also did test single line double component functions (e.g., a + bX +c$X^{\prime}$ [$X^{\prime} \subseteq$ X ; $X^{\prime} \neq$ X] ; a + bY +c$Y^{\prime}$ [$Y^{\prime} \subseteq$ Y ; $Y^{\prime} \neq$ Y] ; a + bZ +c$Z^{\prime}$ [$Z^{\prime} \subseteq$ Z ; $Z^{\prime} \neq$ Z]). We noticed that the use of more than two components in the linear relation did not give statistically better results, so we limited our test functions to a maximum of two components of EWs or ratio of EWs or combination of both. The multi-line functions (two components) that we tested are : a + bX +cY ; a + bX +cZ ; a + bY +cZ . The ratio of the spectral line depths or equivalent widths (R = EW$_{1}$ / EW$_{2}$) are found to be very sensitive temperature indicators (Kovtyukh et al. \citet{Kovtyukh2003}) and tracers of other fundamental parameters with high efficiency and accuracy. So we also checked with all possible combinations of EW ratios of the three lines ( R$_{1}$ = X / Y ; R$_{2}$ = X / Z ; R$_{3}$ = Y / Z). For single-ratio (one component) linear functions we did test : a + R$_{i}$b (i = 1,2,3), and single-line and single-ratio functions we tested : a + bX + cR$_{i}$ ; a + bY + cR$_{i}$ ;  a + bZ + cR$_{i}$ (i =1,2,3) . Like the single-line double component functions, we tried out with single-ratio double component functions : a + bR$_{i}$ + cR$_{i}^{\prime}$ (R$_{i}^{\prime} \subseteq  R_{i}$ ; R$_{i}^{\prime} \neq R_{i}$ ; i = 1,2,3) and also tested with multi-ratio functions : a + bR$_{1}$ + cR$_{2}$ ; a + bR$_{1}$ + cR$_{3}$ ; a + bR$_{2}$ + cR$_{3}$ . \\
$~~~~~$ To find the best-fitting parameters from all possible combinations of linear functional relations, we performed multivariate linear regression and used the adjusted square of the multiple correlation coefficient ($R^{2}_{ap}$ , Rojas-Ayala et al. \citet{RojasAyala2012}), the root mean squared error (RMSE) and the mean absolute deviation (MAD) values to statistically compare the goodness of each fit. We show the calculations of RMSE, MAD and $R^{2}_{ap}$ as equation (\ref{eq_rmse}), equation (\ref{eq_mad}) and equation (\ref{eq_r2ap}) in the appendix section. For a given set of calibration data, the $R^{2}_{ap}$ shows the proportion of variability in a data set that is accounted for by a regression model (an $R^{2}_{ap}$ value closer to 1 means a superior fit) and the RMSE (a lower RMSE means a better fit) evaluates the accuracy and predictive power of the response of the regression model (Khata et al. \citet{Khata2020}). We tested 8001 possible combinations of linear trial functions using the selected EWs and EW ratios and in Table \ref{tab:table4_trial_fun} we show different types of linear trial functions with one example each, alongwith the respective RMSE, MAD and $R^{2}_{ap}$ values for $T_{\rm eff}$, radius and luminosity. In the case of RMSE, the individual errors are squared and the RMSE gives relatively high weight to large errors, whereas for MAD the errors are co-added in a linear fashion attributing equal weightage on small and large errors. This means the RMSE is more suitable when large deviation from the mean is particularly undesirable. Also as our estimated errors on the EWs are expected to be Gaussian in nature, the RMSE is more appropriate to represent a model performance than the MAD. So we selected the best fitting calibration relations for the stellar parameters mainly on the basis of highest $R^{2}_{ap}$ and lowest RMSE values. To get a clear understanding of how much better the chosen relations are than the rest of the trial functions, we show a representation of the RMSE, MAD and $R^{2}_{ap}$ values in Fig. \ref{fig:fig3_trial_fun}. We find that for $T_{\rm eff}$, our chosen function also gives the lowest MAD value alongside the lowest RMSE value, but for the best fitting radius ($MAD_{rad}$ = 0.043) and luminosity ($MAD_{lum}$ = 0.137) relations the MAD values are slightly greater than the lowest estimated values ($MAD_{lowest, rad}$ = 0.040 and $MAD_{lowest, lum}$ = 0.124). The selected calibration relations for the stellar parameters are given as :

\vspace*{-0.3 cm} \begin{equation} \label{eq2}
\begin{split}
\hspace*{0.5 cm} \textbf T_{eff}/K &=\  A + B\times Ca (0.854 \mu m)+C\times \frac{[Mg (1.57 \mu m)]^2}{Ca (0.854 \mu m)}  \\
\end{split}
\end{equation}
\\
\hspace*{0.2cm} \vspace*{0.1 cm}  A = 2738.37 $\pm$ 39.5 ; B = 232.046 $\pm$ 7.6 ; C = 35.462 $\pm$ 2.6 \\
\hspace*{0.8cm} \vspace*{0.1 cm} \footnotesize $R^{2}_{ap}$ = 0.881 ; \hspace*{0.2cm}   RMSE ($T_{\rm eff}$/K) = 99.0 ;    \hspace*{0.2cm} MAD / K = 71.0

\normalsize
\begin{equation} \label{eq3}
\begin{split}
\hspace*{0.6cm} \textbf R/R_\odot\ &=\ A + B\times Ca (0.854 \mu m)+C\times \frac{[Mg (1.57 \mu m)]^2}{Ca (0.866 \mu m)}  \\
\end{split}
\end{equation}
\\
\hspace*{0.0cm} \vspace*{0.1 cm}  A = 0.1102 $\pm$ 0.01 ; B = 0.0757 $\pm$ 0.002 ; C = 0.0231 $\pm$ 0.0009 \\
\hspace*{0.7cm} \footnotesize $R^{2}_{ap}$ = 0.779  ; \hspace*{0.1cm} RMSE ($R/R_{\sun}$) = 0.067 ; \hspace*{0.1cm} MAD / $R_{\sun}$   = 0.043

\vspace*{-0.0cm}
\normalsize
\begin{equation} \label{eq4}
\begin{split}
\hspace*{-0.3 cm}  log \textbf L/L_\odot\ &=\ A + B\times \frac{Ca (0.854 \mu m)}{[Mg (1.57 \mu m)]^{1/2}} +C\times \frac{Mg (1.57 \mu m)}{[Ca (0.866 \mu m)]^{1/2}}  \\
\end{split}
\end{equation}
\\
\hspace*{-0.1cm} \vspace*{0.1 cm}  A = \textendash 3.5063 $\pm$ 0.032  ; B = 0.4896 $\pm$ 0.004  ; C = 0.5363 $\pm$ 0.003 \\
\hspace*{0.4cm} \footnotesize $R^{2}_{ap}$ = 0.798  ; \hspace*{0.1cm} RMSE ($log L/L_{\sun}$) = 0.218 ; \hspace*{0.1cm} MAD / dex = 0.137  \\

\normalsize

\hspace{-0.75 cm} We show the plots of our best-fitting linear relationships and the respective residuals of the parameters for the calibrator stars in Fig. \ref{fig:fig4_parameter_calibration}. The final errors on the stellar parameters are estimated by combining the randomly generated Gaussian error propagated from the EWs of the features with the intrinsic scatter (RMSE) inherent in the calibration relations. In Table \ref{tab:table5_teff_rad_lum_cal}, we present the inferred stellar parameters with associated errors that we estimated using the calibration relations. For our work, we notice that along with the EWs, the simple functional form of the ratios of EWs plays an important role in the calibration. In principle, the change in EWs with $T_{\rm eff}$ for lines with higher excitation potential ($\chi$) should be faster than that for lines with lower '$\chi$' and a ratio of EWs should be temperature sensitive if the excitation potential of the lines differs by as much as possible (Gray 1994). The excitation potential of the Ca II IRT lines are 1.70 eV and 1.69 eV, respectively, and for the Mg I triplet, the '$\chi$' value is around 5.93 eV, so the EW ratios $Mg (1.57 \mu m) / Ca (0.854 \mu m)$ and $Mg (1.57 \mu m) / Ca (0.866 \mu m)$ show better correlation with the stellar parameters than the ratio of the two Ca II IRT lines [i.e., $Ca (0.854 \mu m) / Ca (0.866 \mu m)$]. Also as the EWs show very negligible metallicity dependence within the parameter-space, selecting the lines from different part of the spectra does not hamper the calibration substantially.\\
\hspace*{0.3cm} In Fig. \ref{fig:fig5_radvsteff_lumvsteff_plt}, we show the $T_{\rm eff}$-radius (left panel) and $T_{\rm eff}$-luminosity (right panel) distribution of our sample of the CARMENES M-dwarfs. For comparison, in the left panel, we overplot the radius-temperature relation from Mann et al. (\citet{Mann2015}, hereafter Mann15) and new evolutionary models of ages 0.2 Gyr, 2.0 Gyr, and 10.0 Gyr for pre-main sequence and main sequence low-mass stars down to the hydrogen-burning limit for solar metallicity by Baraffe et al. (\citet{Baraffe2015}, hereafter BHAC15). Our estimated radii are a little larger than what is predicted by Mann15's relation and have a distinct separation from the BHAC15 isochrones. One reason for the disagreement could be that the stellar radii for low-mass stars generally are measured to be larger than model predictions (Boyajian et al. \citet{Boyajian2012}). Another reason could be the metallicity dependence of parameters (Newton et al. \citet{Newton2015}) which is observed in our temperature-radius plane. It can be seen that for $T_{\rm eff}$ < 3700, a large part of our sample is metal poor. In the right panel of Fig. \ref{fig:fig5_radvsteff_lumvsteff_plt}, we also show the overplot of the BHAC15 isochrones and the 5 Gyr, Dartmouth isochrones (Dotter et al. \citet{Dotter2008}; Feiden et al. \citet{Feiden2011}) for super-solar ([Fe/H] = +0.30 dex), solar ([Fe/H] = 0.0 dex) and sub-solar ([Fe/H] = -0.50 dex) metallicity. Here also metallicity has a significant effect on the models (as noted by Newton et al. \citet{Newton2015}), and the Dartmouth models predict that for cool stars, there will be a difference of 1 in absolute magnitude for a star with [Fe/H] = -0.50 dex, compared to another star of same effective temperature with [Fe/H] = +0.30 dex.\\
$~~~~~$ In Table \ref{tab:tableA1_estimated_para}, we present the estimated EWs of the selected spectral features that we use for the calibration and the stellar parameters of the CARMENES M-dwarf stars. We also show a plot of the distribution of the stellar parameters of our M-dwarf sample in Fig. \ref{fig:fig6_para_dist} with the parameter values ranging $T_{\rm eff}$ (2849 < $T_{\rm eff}$(K) < 4206 ), radius (0.141 < $R/R_{\sun}$ < 0.776) and luminosity (\textendash 3.00 < log $L/L_{\sun}$ < \textendash 0.643).

\subsection{Comparison with literature studies}

\begin{figure}
\centering
\vspace*{-0.4 cm}
\begin{tabular}{ccc}
\hspace*{-0.8 cm} \vspace*{-0.1 cm}
\includegraphics[width= 3.7 in, height = 2.7 in, clip]{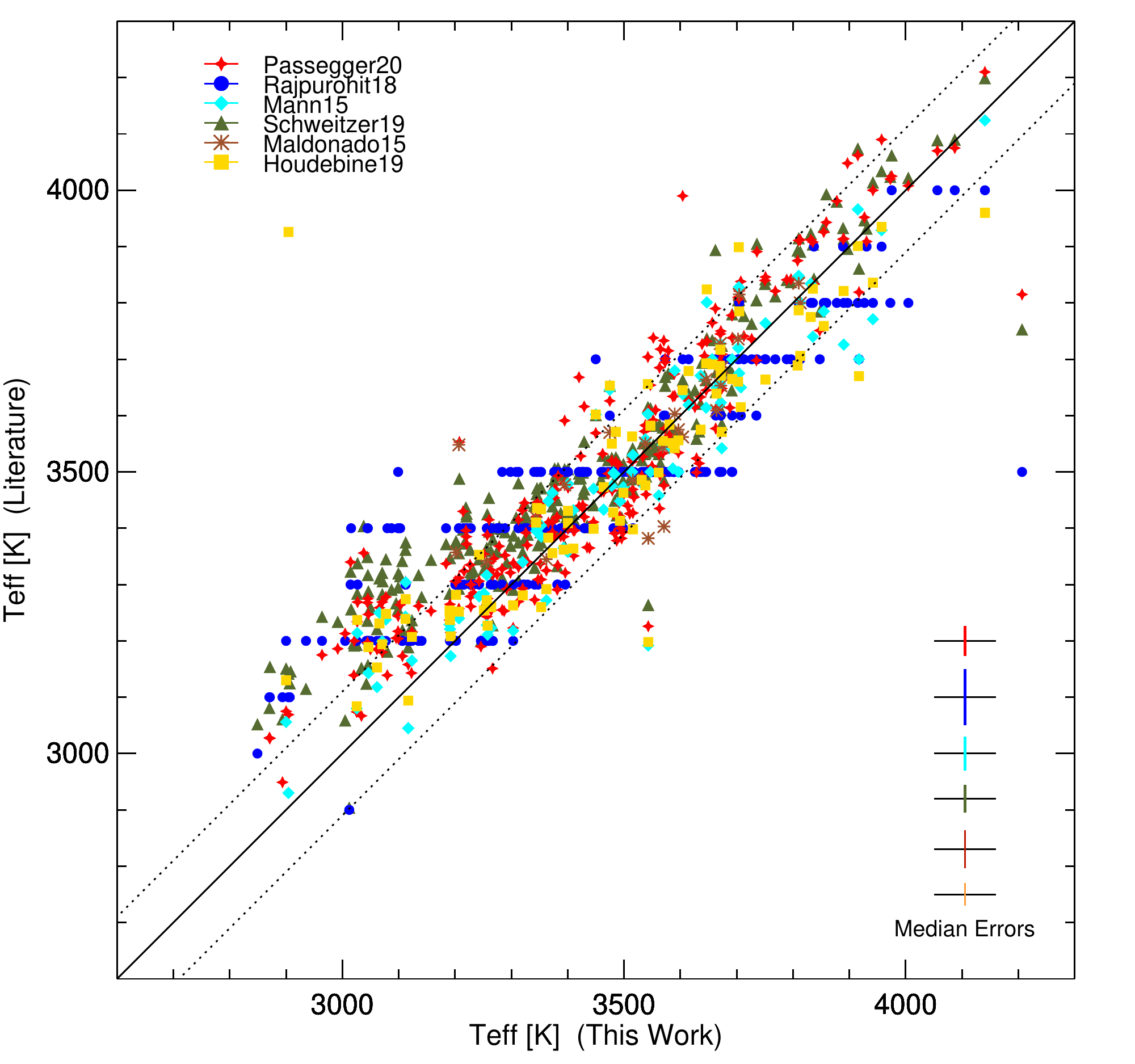} \\ 
\hspace*{-0.8 cm} \vspace*{-0.1 cm} \includegraphics[width= 3.7 in, height = 2.7 in, clip]{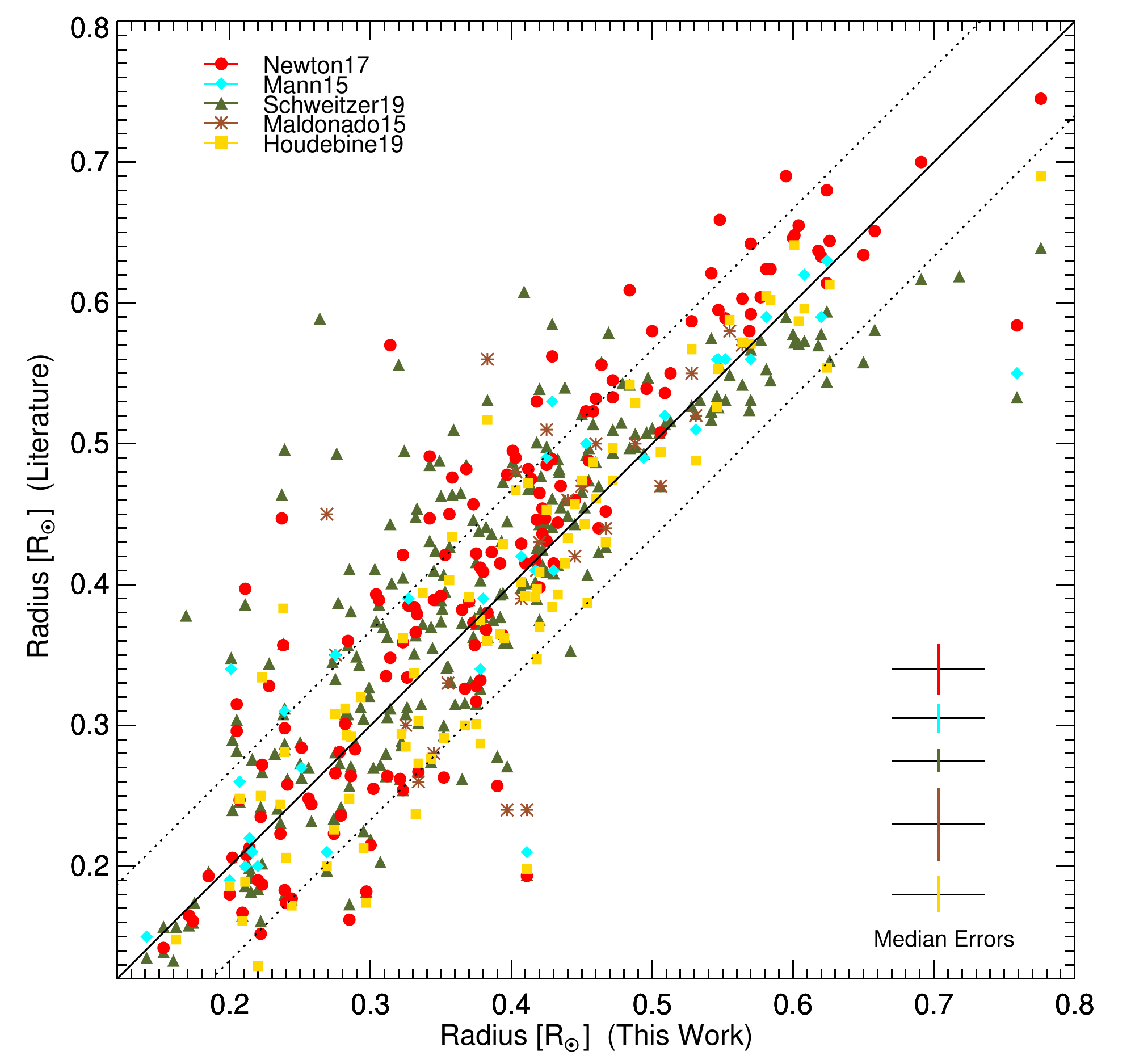}\\
\hspace*{-0.8 cm} \includegraphics[width= 3.7 in, height = 2.7 in, clip]{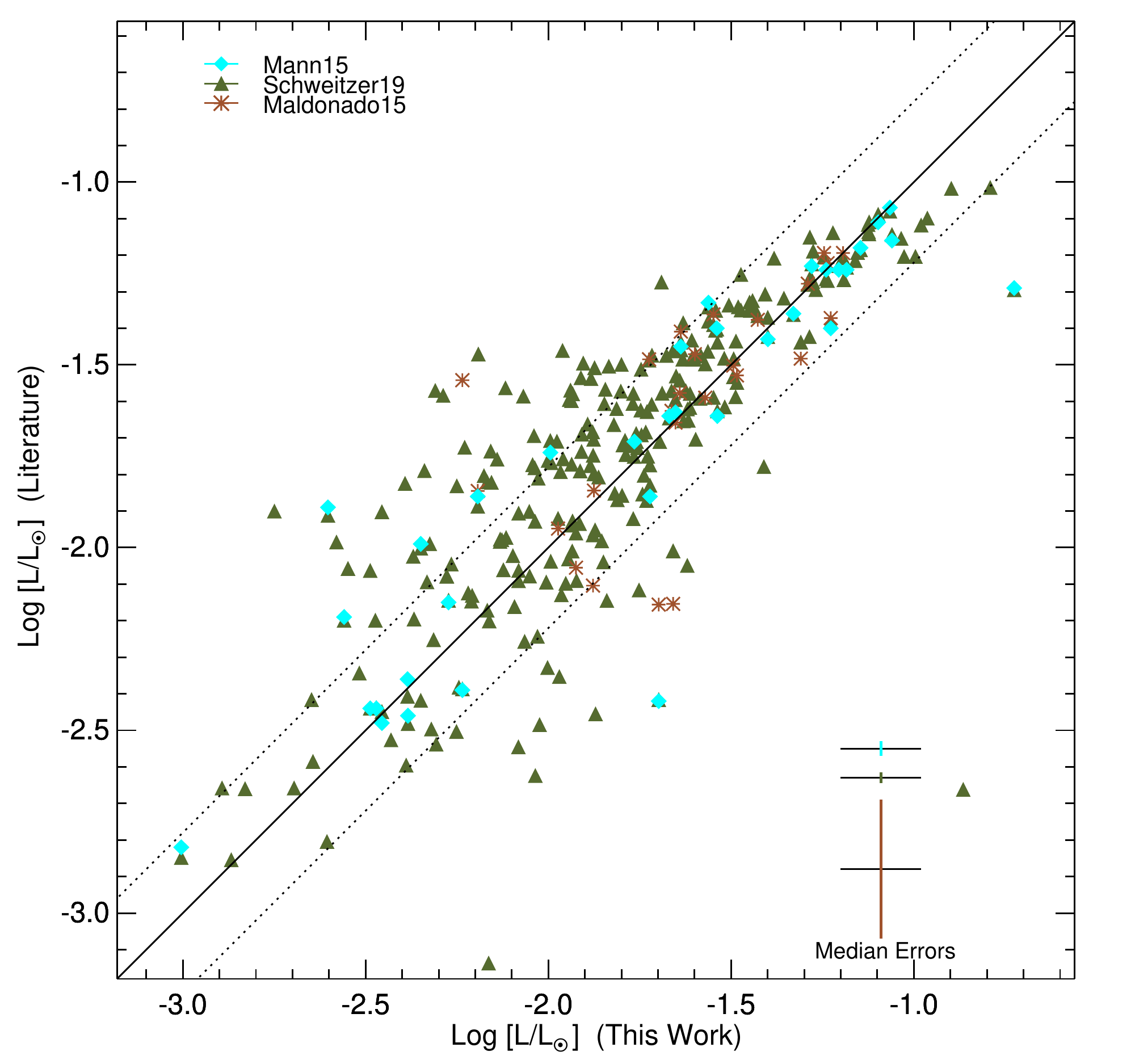}
\end{tabular}
\caption{Comparisons of $T_{\rm eff}$ (top panel), radius (middle panel) and luminosity (bottom panel) with literature results (vertical axes) and this work (horizontal axes). The 1:1 relation is indicated by the solid black lines and the dotted lines indicate the typical size of the error bars associated with our work. The median errors associated with the literature values are shown in the lower right corner of each plot. Different determination methods of literature values are indicated by different symbols and colors. Legends used in the plots : Passegger20 (Passegger et al. \citet{Passegger2020}); Rajpurohit18 (Rajpurohit et al. \citet{Rajpurohit2018}); Mann15 (Mann et al. \citet{Mann2015}); Schweitzer19 (Schweitzer et al. \citet{Schweitzer2019}); Maldonado15 (Maldonado et al. \citet{Maldonado2015}); Houdebine19 (Houdebine et al. \citet{Houdebine2019}); Newton17 (Newton et al. \citet{Newton2017}).}
\label{fig:fig7_para_comparison}
\end{figure}

In Fig. \ref{fig:fig7_para_comparison}, we show the comparison of the results that we get in this work (horizontal axes) with the literature parameter values obtained in other works (vertical axes) using different techniques for the overlapping M-dwarfs. In Table \ref{tab:table6_medianstddev}, we present the median and standard deviation of the differences between the various literature values and the values determined here. For better readability, we do not use error bars in the plots; rather, we show median errors of all the comparisons in the lower right side of each plot.\\
$~~~~$In the top panel of Fig. \ref{fig:fig7_para_comparison}, we compare our estimated $T_{\rm eff}$ with different literature works. Passegger et al. (\citet{Passegger2020}, hereafter Passegger20) derived $T_{\rm eff}$ using neural network architecture for CARMENES spectra, Mann15 calculated $T_{\rm eff}$ by comparing their optical spectra with BT-Settl (PHOENIX) models, Maldonado et al. (\citet{Maldonado2015}, hereafter Maldonado15) used ratios of pseudo EWs of spectral features as temperature diagnostic and Houdebine et al. (\citet{Houdebine2019}, hereafter Houdebine19) used a photometric approach and correlated (R - I)$_{C}$ color with $T_{\rm eff}$. Apart from some minor scatter, all these results show reasonably good median and standard deviation values of differences with our estimation and overlap (within their error) of the dotted line region which represents $\pm$ 110K from a one-to-one comparison. Comparing the CARMENES spectra with the BT-Settl models both in the optical and in the NIR simultaneously, Rajpurohit et al. (\citet{Rajpurohit2018}, hereafter Rajpurohit18) estimated $T_{\rm eff}$ with a median error around 100K and Schweitzer et al. (\citet{Schweitzer2019}, hereafter Schweitzer19) measured the $T_{\rm eff}$ by fitting the CARMENES spectra of the visual (VIS) channel with PHOENIX-ACES synthetic spectra (Husser et al. \citet{Husser2013}). Rajpurohit18 and Schweitzer19 show relatively consistent results as the data points stay well within the dotted line region down to 3500K. Below 3500K some of the estimated $T_{\rm eff}$ values (from Rajpurohit18 and Schweitzer19) are up to 300K higher than our values. Although for both the cases, the overall median and standard deviation in differences are less than 90K.\\
$~~~~$ The comparison of the radius is shown in the middle panel of Fig. \ref{fig:fig7_para_comparison}. Newton et al. (\citet{Newton2017}, hereafter Newton17) and Mann15 empirically calibrated stellar radius, Maldonado15 used photometric estimates of radius to calibrate empirical relations using temperatures and metallicities, and Houdebine19 used V-band absolute magnitude and $T_{\rm eff}$ to determine the radii. We observe that the radii estimated by Newton17 and Houdebine19 are larger than our values down to 0.3R$_{\sun}$ and for radii smaller than that, their values are slightly lower than ours. Schweitzer19 has the biggest scatter (compared to our results), with radii that are generally slightly larger than ours. These discrepancies might be related to the activity level of M-dwarfs (Morales, Ribas and Jordi \citet{Morales2008}) as either strong magnetic fields predictably inhibiting convection or flux conservation in a magnetic spot-covered stellar surface (L\'opez-Morales $\&$ Ribas \citet{Lopez2005}) causes larger radii for a given $T_{\rm eff}$. Also, Schweitzer19 used effective temperature to calculate the radii using Stefan-Boltzmann's law while they derived their $T_{\rm eff}$ by fitting with synthetic spectra, and experimental evidence suggests that theoretical models have limitations to describe star's $T_{\rm eff}$ accurately for a value lower than 4000K and that could add errors in radii. However, despite the scatter, the literature radii show consistency with our estimations with the median and standard deviation of differences being well below the median value of our radius uncertainty (0.067 R$_{\sun}$).\\ 
$~~~~$ In the bottom panel of Fig. \ref{fig:fig7_para_comparison}, we show the luminosity comparison for three literature results: Mann15 estimated luminosity using empirical relations, using apparent magnitudes and distances, Schweitzer19 photometrically determined luminosity, and Maldonado15 used the Stefan-Boltzmann law. In general Schweitzer19 is in good agreement with the one-to-one relationship, however there is some scatter in the luminosity (log L/$L_{\sun}$) in the range of -1.5 to -2.0 dex. This mismatch could arise from the discrepancy in the value of the apparent magnitudes as experimental evidence suggests that active and inactive M dwarfs define different sequences in the luminosity-color plane (Stauffer $\&$ Hartmann \citet{Stauffer1986}) plane and the uncertainty of distance calculation adds to the inconsistency.

\begin{table}
\hspace{-2.5 cm}
 \centering
 \caption{ Median and standard deviation (Std. Dev.) of differences between the results of literature and this work for stellar parameters}
\hspace*{-0.3 cm} \begin{tabular}{lccc}
  \hline \hline
      & & Median / Std. Dev.   & \\
  Literature &  $T_{\rm eff}$ (K) & \hspace{-0.2 cm} Radius ($R_{\sun}$) & \hspace{-0.2 cm} log (L/$L_{\sun}$) \\
  \hline
 Schweitzer et al. (\citet{Schweitzer2019})  & \hspace{-0.2 cm} 75 / 80   & \hspace{-0.2 cm} 0.038 / 0.050  & \hspace{-0.2 cm} 0.13 / 0.22 \\
 Maldonado et al. (\citet{Maldonado2015})  & \hspace{-0.2 cm} 33 / 75   & \hspace{-0.2 cm} 0.025 / 0.056  & \hspace{-0.2 cm} 0.064 / 0.179 \\
  Mann et al. (\citet{Mann2015})  & \hspace{-0.2 cm} 45 / 60   & \hspace{-0.2 cm} 0.019 / 0.052  & \hspace{-0.2 cm} 0.10 / 0.19 \\
 Houdebine et al. (\citet{Houdebine2019})  & \hspace{-0.2 cm} 58 / 63   & \hspace{-0.2 cm} 0.012 / 0.055  & \hspace{-0.2 cm} \textendash \\
 Passegger et al. (\citet{Passegger2020})  & \hspace{-0.2 cm} 73 / 69   & \hspace{-0.2 cm} \textendash  & \hspace{-0.2 cm} \textendash \\
 Rajpurohit et al. (\citet{Rajpurohit2018})  & \hspace{-0.2 cm} 83 / 86   & \hspace{-0.2 cm} \textendash  & \hspace{-0.2 cm} \textendash \\
 Newton et al. (\citet{Newton2017})  & \hspace{-0.2 cm} \textendash   & \hspace{-0.2 cm} 0.043 / 0.045  & \hspace{-0.2 cm} \textendash \\

   \hline
 \end{tabular}
    \label{tab:table6_medianstddev}
\end{table}

\section{Summary and Conclusion}

We have estimated effective temperature, radius, and luminosity (L/${L}_{\odot})$ with individual error bars for a sample of 271 M-dwarf stars with spectral types (M0V-M7V) observed in the CARMENES radial-velocity planet survey. We establish new empirical calibration relationships to determine fundamental parameters of low mass M-dwarf stars using EWs and EW ratios of spectral absorption features taken simultaneously from high resolution (R$\sim90000$) optical (0.52\textendash0.96 $\mu$m) and near-infrared (0.96\textendash 1.71 $\mu$m) spectra. The estimation of our physical properties is mostly model-independent, and the main results are summarized as follow:

\begin{enumerate}

\item[(1)] The optical and NIR spectra of CARMENES M-dwarfs contain several important spectral absorption features, which show a legitimate strong correlation with the physical parameters. We select a sample of 23 stars with interferometrically measured radii as our calibrators. We perform principal component analysis (PCA), looking for the strongest correlations among the measured EWs of our calibration sample and derive the linear Pearson correlation coefficient to identify the most suitable tracers of the stellar parameters.
\item [(2)] The optical (Ca II at 0.854 $\mu$m, Ca II at 0.866 $\mu$m) and NIR (Mg I at 1.574 $\mu$m) features show a strong correlation with the stellar parameters giving the best correlation coefficient values. We find that the absorption lines of low and high excitation potentials ($\chi$) respond differently to the change of stellar parameters and the inclusion of the ratio of the equivalent widths with different excitation potentials in the calibration relations gives comparatively superior results.
\item [(3)] We investigate simple parameterizations of two EWs or EW ratios or a combination of both and examine every possible single line or multiline linear function. We have performed multivariate linear regression to determine the best fitting parameters and use the adjusted square of the multiple correlation coefficient ($R^{2}_{ap}$), the RMSE, and the mean absolute deviation (MAD) values to statistically compare the goodness of each fit.   
\item [(4)] We notice that the simultaneous use of the EWs and ratios of EWs of these selected features from both the optical and NIR band give the best fit linear calibration relationships of $T_{\rm eff}$, radius, and luminosity with the RMSE values of 99K, 0.06${R}_{\odot}$ and 0.22 dex respectively. Our parameter values cover ranges in the parameter-space of (2849 < $T_{\rm eff}$(K) < 4206 ), (0.141 < $R/R_{\sun}$ < 0.776) and (\textendash 3.00 < log $L/L_{\sun}$ < \textendash 0.643). The radii and luminosities derived in this work are found to be larger than what is predicted by the theoretical evolutionary models and show some metallicity dependence as we see for $T_{\rm eff}$ < 3700 K, where our sample is mostly metal poor.
\item [(5)] We also explore and compare our results with literature values obtained with other methods where available. Though the comparison plots display little differences in terms of the median and standard deviation values, the overall agreement is good. Our calibrations can be very useful for understanding the properties of potential planet-host candidates and distant cool stars without parallaxes and flux calibrated spectra.

\end{enumerate}

\section{Acknowledgements}

The authors are grateful to the reviewer for all the valuable comments and suggestions, which helped to improve the overall quality of this paper. The work is supported by S. N. Bose National Centre for Basic Sciences under the Department of Science and Technology (DST), Govt. of India. The research work leading to this publication has received funding from the INSPIRE Fellowship scheme by the DST, Govt. of India. The result of this research is "Based on data from the CARMENES data archive at CAB (INTA-CSIC)." The authors acknowledge the use of the SIMBAD and VizieR database, operated at CDS, Strasbourg, France. This research made use of the IDL-based tellrv and nirew packages developed by E. R. Newton.

\section{Data Availability}

All the observational data that we have used for the research work of this paper are publicly available in the CARMENES data archive at CAB (INTA-CSIC) (\url{http://carmenes.cab.inta-csic.es/gto/jsp/reinersetal2018.jsp}). Table \ref{tab:table4_trial_fun} is available in its entirety as online supplementary material in the journal.


\appendix
\section{Calculations of RMSE, MAD and $R^{2}_{ap}$}

The root mean squared error (RMSE) is the square root of the variance of the residuals. The RMSE is defined as :

\vspace*{-0.3 cm} \begin{equation} \label{eq_rmse}
 \hspace*{2.0 cm} RMSE = \sqrt{ \sum\limits_{i=1}^n \frac{(\widehat{y_{i}} - y_{i})^2}{(n - p)} } \\
\end{equation}

\hspace{-0.75 cm} where, n is the number of data points (i.e., number of calibrators), p is the number of predictors used in the model, $y_{i}$ is the value of the response variable i, and $\widehat{y_{i}}$ is the value predicted by the regression model for given $y_{i}$.

The mean absolute deviation (MAD) measures the statistical dispersion or variability in a dataset. The MAD is defined as : 

\vspace*{-0.3 cm} \begin{equation} \label{eq_mad}
 \hspace*{2.5 cm} MAD = \sum\limits_{i=1}^n \frac{|y_{i} - \bar{y}|}{n}  \\
\end{equation}

\hspace{-0.75 cm} where, $\bar{y}$ is the overall mean of the observations $y_{i}$.

The adjusted square of the multiple correlation coefficient ($R^{2}_{ap}$) represents the proportion of variability in a data set that is accounted for by a regression model. The $R^{2}_{ap}$ is defined as :

\vspace*{-0.3 cm} \begin{equation} \label{eq_r2ap}
 \hspace*{2.0 cm} R^{2}_{ap} = 1 - \frac{ (n - 1) \sum\limits_{i=1}^n (\widehat{y_{i}} - y_{i})^2 }{ (n - p)\sum\limits_{i=1}^n (y_{i} - \bar{y_{i}})^2} \\
\end{equation}

\section{TABLES}

\label{sec:abbreviations}

\begin{table*}
 \centering
 \caption{EWs of the selected spectral features and estimated stellar parameters of the CARMENES M-dwarf stars.}
  \label{tab:tableA1_estimated_para}
 \begin{tabular}{lcccccccccccc}
  \hline \hline

\hline \multicolumn{1}{c}{Karmn$^{(a)}$} & \multicolumn{1}{c}{Name} & \multicolumn{1}{c}{Sp.type$^{(b)}$} & \multicolumn{1}{c}{Ca I EW (${\AA}$)} & \multicolumn{1}{c}{Ca I EW (${\AA}$)} & \multicolumn{1}{c}{Mg I EW (${\AA}$)} & \multicolumn{1}{c}{$T_{eff}^{(c)}$} & \multicolumn{1}{c}{Radius$^{(c)}$} & \multicolumn{1}{c}{Luminosity$^{(c)}$} & \\ 

\multicolumn{1}{c}{} & \multicolumn{1}{c}{} & \multicolumn{1}{c}{} & \multicolumn{1}{c}{(0.854 $\mu$m)} & \multicolumn{1}{c}{(0.866 $\mu$m)} & \multicolumn{1}{c}{(1.57 $\mu$m)} & \multicolumn{1}{c}{(K)} & \multicolumn{1}{c}{($R_{\sun}$)} & \multicolumn{1}{c}{(log $L/L_{\sun}$)} & \\ 
\hline

J00051+457 &GJ 2          	 &M1.0 V&3.491$\pm$0.608&2.426$\pm$0.375&3.761$\pm$0.193&3692$\pm$107&0.509$\pm$0.067&\textendash1.330$\pm$0.224& \\
J00067-075 &GJ 1002       	 &M5.5 V&0.577$\pm$0.555&1.368$\pm$0.355&0.590$\pm$0.130&2894$\pm$129&0.160$\pm$0.069&\textendash2.868$\pm$0.232& \\
J00162+198E&GJ 1006B      	 &M4.0 V&1.022$\pm$0.505&1.452$\pm$0.311&1.060$\pm$0.294&3015$\pm$108&0.205$\pm$0.067&\textendash2.548$\pm$0.224& \\
J00183+440 &GJ 15A        	 &M1.0 V&3.253$\pm$0.430&2.926$\pm$0.275&2.123$\pm$0.218&3542$\pm$104&0.392$\pm$0.066&\textendash1.747$\pm$0.222& \\
J00184+440 &GJ 15B        	 &M3.5 V&2.400$\pm$0.523&2.272$\pm$0.320&0.730$\pm$0.192&3303$\pm$110&0.297$\pm$0.067&\textendash1.871$\pm$0.226& \\
J00286-066 &GJ 1012       	 &M4.0 V&1.750$\pm$0.598&1.824$\pm$0.394&1.856$\pm$0.353&3214$\pm$128&0.286$\pm$0.071&\textendash2.140$\pm$0.241& \\
J00389+306 &GJ 26         	 &M2.5 V&3.145$\pm$0.297&2.314$\pm$0.205&2.723$\pm$0.142&3552$\pm$110&0.422$\pm$0.067&\textendash1.613$\pm$0.226& \\
J00570+450 &G 172-030            &M3.0 V&1.908$\pm$0.511&2.634$\pm$0.321&1.458$\pm$0.138&3221$\pm$141&0.273$\pm$0.072&\textendash2.251$\pm$0.250& \\
J01013+613 &GJ 47         	 &M2.0 V&2.826$\pm$0.479&2.237$\pm$0.318&2.326$\pm$0.087&3462$\pm$119&0.380$\pm$0.068&\textendash1.765$\pm$0.232& \\
J01025+716 &GJ 48         	 &M3.0 V&2.500$\pm$0.259&2.417$\pm$0.190&2.676$\pm$0.245&3420$\pm$103&0.368$\pm$0.066&\textendash1.835$\pm$0.221& \\
J01026+623 &GJ 49         	 &M1.5 V&3.699$\pm$0.251&3.031$\pm$0.176&3.797$\pm$0.057&3735$\pm$104&0.500$\pm$0.066&\textendash1.407$\pm$0.221& \\
J01048-181 &GJ 1028       	 &M5.0 V&0.459$\pm$0.728&0.965$\pm$0.461&0.585$\pm$0.200&2871$\pm$126&0.153$\pm$0.067&\textendash2.893$\pm$0.223& \\
J01125-169 &GJ 54.1       	 &M4.5 V&0.662$\pm$0.690&1.780$\pm$0.355&0.388$\pm$0.117&2900$\pm$108&0.162$\pm$0.067&\textendash2.830$\pm$0.224& \\
J01339-176 &LP 768-113     	 &M4.0 V&1.150$\pm$0.802&1.534$\pm$0.458&1.128$\pm$0.226&3044$\pm$118&0.216$\pm$0.069&\textendash2.487$\pm$0.233& \\
J01433+043 &GJ 70         	 &M2.0 V&3.167$\pm$0.578&2.732$\pm$0.302&2.809$\pm$0.219&3562$\pm$111&0.417$\pm$0.068&\textendash1.669$\pm$0.227& \\
J01518+644 &GJ 3117A      	 &M2.5 V&2.854$\pm$0.576&2.402$\pm$0.306&3.354$\pm$0.386&3540$\pm$122&0.434$\pm$0.071&\textendash1.582$\pm$0.236& \\
J02002+130 &GJ 83.1       	 &M3.5 V&1.201$\pm$0.599&0.842$\pm$0.306&0.532$\pm$0.211&3025$\pm$108&0.209$\pm$0.067&\textendash2.389$\pm$0.224& \\
J02015+637 &GJ 3126       	 &M3.0 V&2.256$\pm$0.522&2.050$\pm$0.297&3.204$\pm$0.200&3423$\pm$120&0.397$\pm$0.070&\textendash1.689$\pm$0.239& \\
J02070+496 &G 173-037            &M3.5 V&3.153$\pm$0.572&2.440$\pm$0.313&1.514$\pm$0.179&3496$\pm$109&0.371$\pm$0.067&\textendash1.732$\pm$0.225& \\
J02088+494 &GJ 3136       	 &M3.5 V&1.714$\pm$0.432&1.365$\pm$0.312&2.110$\pm$0.218&3228$\pm$114&0.315$\pm$0.068&\textendash1.960$\pm$0.227& \\
J02123+035 &GJ 87         	 &M1.5 V&3.211$\pm$0.254&2.742$\pm$0.202&3.305$\pm$0.139&3604$\pm$106&0.445$\pm$0.067&\textendash1.571$\pm$0.223& \\
J02222+478 &GJ 96         	 &M0.5 V&3.763$\pm$0.197&3.150$\pm$0.191&5.078$\pm$0.116&3855$\pm$107&0.584$\pm$0.068&\textendash1.154$\pm$0.225& \\
J02336+249 &GJ 102        	 &M4.0 V&1.283$\pm$0.577&1.008$\pm$0.362&1.121$\pm$0.247&3071$\pm$124&0.236$\pm$0.068&\textendash2.314$\pm$0.228& \\
J02358+202 &GJ 104        	 &M2.0 V&3.621$\pm$0.455&3.036$\pm$0.314&3.113$\pm$0.099&3674$\pm$112&0.458$\pm$0.069&\textendash1.543$\pm$0.229& \\
J02362+068 &GJ 105B       	 &M4.0 V&2.124$\pm$0.488&1.859$\pm$0.330&1.098$\pm$0.067&3251$\pm$114&0.286$\pm$0.068&\textendash2.082$\pm$0.228& \\
J02442+255 &GJ 109        	 &M3.0 V&2.495$\pm$0.221&2.445$\pm$0.224&1.579$\pm$0.093&3353$\pm$110&0.323$\pm$0.067&\textendash1.992$\pm$0.226& \\
J02519+224 &RBS 365      	 &M4.0 V&1.598$\pm$0.348&1.129$\pm$0.381&1.475$\pm$0.135&3157$\pm$129&0.276$\pm$0.070&\textendash2.117$\pm$0.236& \\
J02530+168 &Teegarden's Star     &M7.0 V&1.174$\pm$0.426&1.198$\pm$0.371&0.216$\pm$0.102&3012$\pm$130&0.200$\pm$0.071&\textendash2.163$\pm$0.242& \\
J02565+554W&GJ 119A       	 &M1.0 V&4.133$\pm$0.461&3.128$\pm$0.388&4.819$\pm$0.303&3897$\pm$103&0.595$\pm$0.067&\textendash1.123$\pm$0.221& \\
J03181+382 &GJ 134        	 &M1.5 V&4.004$\pm$0.214&3.057$\pm$0.182&5.309$\pm$0.293&3917$\pm$115&0.626$\pm$0.069&\textendash1.027$\pm$0.231& \\
J03213+799 &GJ 133        	 &M2.0 V&2.988$\pm$0.368&2.455$\pm$0.317&3.075$\pm$0.149&3544$\pm$106&0.425$\pm$0.067&\textendash1.619$\pm$0.223& \\
J03217-066 &GJ 3218       	 &M2.0 V&3.274$\pm$0.354&2.536$\pm$0.312&2.626$\pm$0.139&3573$\pm$108&0.421$\pm$0.068&\textendash1.632$\pm$0.224& \\
J03463+262 &GJ 154        	 &M0.0 V&4.141$\pm$0.439&3.458$\pm$0.398&4.320$\pm$0.186&3859$\pm$119&0.548$\pm$0.069&\textendash1.285$\pm$0.232& \\
J03473-019 &G 080-021            &M3.0 V&1.902$\pm$0.378&1.648$\pm$0.308&2.736$\pm$0.286&3319$\pm$120&0.359$\pm$0.068&\textendash1.800$\pm$0.229& \\
J03531+625 &Ross 567           	 &M3.0 V&1.115$\pm$0.231&0.618$\pm$0.151&2.573$\pm$0.070&3208$\pm$119&0.442$\pm$0.066&\textendash1.410$\pm$0.221& \\
J04225+105 &LSPM J0422+1031      &M3.5 V&2.184$\pm$0.467&1.958$\pm$0.297&2.195$\pm$0.262&3323$\pm$119&0.332$\pm$0.067&\textendash1.943$\pm$0.227& \\
J04290+219 &GJ 169        	 &M0.5 V&4.538$\pm$0.130&3.200$\pm$0.090&6.685$\pm$0.101&4141$\pm$105&0.776$\pm$0.066&\textendash0.642$\pm$0.221& \\
J04376+528 &GJ 172        	 &M0.0 V&4.393$\pm$0.310&3.530$\pm$0.202&4.971$\pm$0.246&3957$\pm$108&0.604$\pm$0.067&\textendash1.122$\pm$0.225& \\
J04376-110 &GJ 173        	 &M1.5 V&3.528$\pm$0.171&2.986$\pm$0.103&2.801$\pm$0.125&3636$\pm$107&0.438$\pm$0.067&\textendash1.605$\pm$0.223& \\
J04429+189 &GJ 176        	 &M2.0 V&3.325$\pm$0.240&2.757$\pm$0.141&2.740$\pm$0.261&3590$\pm$106&0.425$\pm$0.067&\textendash1.637$\pm$0.223& \\
J04429+214 &2M J04425586+21282305&M3.5 V&1.816$\pm$0.441&2.403$\pm$0.303&1.746$\pm$0.120&3219$\pm$117&0.277$\pm$0.068&\textendash2.229$\pm$0.228& \\
J04472+206 &RX J0447.2+2038      &M5.0 V&0.610$\pm$0.842&0.879$\pm$0.472&0.697$\pm$0.112&2908$\pm$107&0.169$\pm$0.066&\textendash2.749$\pm$0.222& \\
J04520+064 &GJ 179        	 &M3.5 V&2.427$\pm$0.470&1.615$\pm$0.299&2.353$\pm$0.301&3382$\pm$111&0.373$\pm$0.068&\textendash1.738$\pm$0.227& \\
J04538-177 &GJ 180        	 &M2.0 V&3.077$\pm$0.276&2.435$\pm$0.173&3.413$\pm$0.186&3587$\pm$113&0.454$\pm$0.068&\textendash1.518$\pm$0.225& \\
J04588+498 &GJ 181        	 &M0.0 V&4.242$\pm$0.157&3.433$\pm$0.103&5.808$\pm$0.114&4005$\pm$112&0.658$\pm$0.068&\textendash0.963$\pm$0.225& \\
J05019+011 &1RXS J050156.7+010845 &M4.0 V&1.340$\pm$0.430&1.228$\pm$0.252&1.665$\pm$0.178&3123$\pm$124&0.264$\pm$0.067&\textendash2.192$\pm$0.223& \\
J05019-069 &GJ 3323       	 &M4.0 V&1.284$\pm$0.476&2.081$\pm$0.314&0.598$\pm$0.108&3046$\pm$112&0.211$\pm$0.068&\textendash2.471$\pm$0.227& \\
J05033-173 &GJ 3325       	 &M3.0 V&2.677$\pm$0.440&1.906$\pm$0.281&1.582$\pm$0.287&3393$\pm$124&0.343$\pm$0.072&\textendash1.849$\pm$0.253& \\
J05062+046 &RX J0506.2+0439	 &M4.0 V&1.447$\pm$0.499&1.217$\pm$0.289&1.015$\pm$0.170&3099$\pm$110&0.239$\pm$0.067&\textendash2.309$\pm$0.226& \\
J05127+196 &GJ 192        	 &M2.0 V&3.151$\pm$0.416&2.420$\pm$0.269&2.771$\pm$0.196&3556$\pm$109&0.422$\pm$0.068&\textendash1.624$\pm$0.228& \\
J05280+096 &GJ 203        	 &M3.5 V&1.980$\pm$0.445&2.024$\pm$0.291&1.271$\pm$0.397&3227$\pm$112&0.279$\pm$0.068&\textendash2.167$\pm$0.228& \\
J05314-036 &GJ 205        	 &M1.5 V&3.656$\pm$0.174&3.209$\pm$0.121&4.829$\pm$0.194&3813$\pm$105&0.555$\pm$0.067&\textendash1.246$\pm$0.223& \\
J05337+019 &GJ 207.1      	 &M2.5 V&1.948$\pm$0.429&1.775$\pm$0.259&2.547$\pm$0.418&3308$\pm$122&0.342$\pm$0.069&\textendash1.883$\pm$0.232& \\
J05348+138 &GJ 3356       	 &M3.5 V&1.730$\pm$0.425&2.127$\pm$0.265&1.979$\pm$0.198&3220$\pm$114&0.284$\pm$0.068&\textendash2.176$\pm$0.229& \\
J05360-076 &GJ 3357       	 &M4.0 V&2.124$\pm$0.503&1.746$\pm$0.280&1.465$\pm$0.247&3267$\pm$116&0.299$\pm$0.069&\textendash2.052$\pm$0.231& \\
J05365+113 &GJ 208        	 &M0.0 V&4.006$\pm$0.145&3.320$\pm$0.107&5.283$\pm$0.127&3915$\pm$106&0.608$\pm$0.068&\textendash1.098$\pm$0.225& \\
J05366+112 &2M J05363846+1117487 &M4.0 V&1.347$\pm$0.521&1.208$\pm$0.277&1.400$\pm$0.209&3103$\pm$117&0.250$\pm$0.068&\textendash2.265$\pm$0.230& \\
J05421+124 &GJ 213        	 &M4.0 V&1.240$\pm$0.482&1.779$\pm$0.271&1.176$\pm$0.247&3066$\pm$110&0.222$\pm$0.067&\textendash2.473$\pm$0.225& \\
J06000+027 &GJ 3379       	 &M4.0 V&1.100$\pm$0.413&1.732$\pm$0.241&1.008$\pm$0.146&3026$\pm$106&0.207$\pm$0.067&\textendash2.559$\pm$0.223& \\
J06011+595 &GJ 3378       	 &M3.5 V&2.171$\pm$0.467&1.840$\pm$0.247&2.185$\pm$0.448&3320$\pm$109&0.334$\pm$0.068&\textendash1.923$\pm$0.225& \\
J06103+821 &GJ 226        	 &M2.0 V&2.873$\pm$0.239&2.620$\pm$0.162&2.324$\pm$0.137&3472$\pm$103&0.375$\pm$0.066&\textendash1.813$\pm$0.221& \\
J06105-218 &GJ 229A       	 &M0.5 V&3.776$\pm$0.216&3.063$\pm$0.120&4.399$\pm$0.137&3796$\pm$103&0.542$\pm$0.066&\textendash1.276$\pm$0.221& \\
J06246+234 &GJ 232        	 &M4.0 V&1.642$\pm$1.132&0.796$\pm$0.540&0.433$\pm$0.623&3123$\pm$122&0.240$\pm$0.069&\textendash2.024$\pm$0.236& \\
J06371+175 &GJ 239        	 &M0.0 V&3.547$\pm$0.285&2.707$\pm$0.170&2.924$\pm$0.113&3647$\pm$106&0.452$\pm$0.067&\textendash1.537$\pm$0.222& \\
J06396-210 &LP 780-032         	 &M4.0 V&2.510$\pm$0.471&1.523$\pm$0.208&2.072$\pm$0.191&3381$\pm$118&0.365$\pm$0.068&\textendash1.752$\pm$0.230& \\

\hline
\end{tabular}
     \raggedright \hspace*{14.5 cm} (continued on next page ..) \\
\end{table*}

\begin{table*}
 \centering
 \contcaption{from the previous page.}
 \begin{tabular}{lcccccccccccc}
  \hline \hline

\hline \multicolumn{1}{c}{Karmn$^{(a)}$} & \multicolumn{1}{c}{Name} & \multicolumn{1}{c}{Sp.type$^{(b)}$} & \multicolumn{1}{c}{Ca I EW (${\AA}$)} & \multicolumn{1}{c}{Ca I EW (${\AA}$)} & \multicolumn{1}{c}{Mg I EW (${\AA}$)} & \multicolumn{1}{c}{$T_{eff}^{(c)}$} & \multicolumn{1}{c}{Radius$^{(c)}$} & \multicolumn{1}{c}{Luminosity$^{(c)}$} & \\ 

\multicolumn{1}{c}{} & \multicolumn{1}{c}{} & \multicolumn{1}{c}{} & \multicolumn{1}{c}{(0.854 $\mu$m)} & \multicolumn{1}{c}{(0.866 $\mu$m)} & \multicolumn{1}{c}{(1.57 $\mu$m)} & \multicolumn{1}{c}{(K)} & \multicolumn{1}{c}{($R_{\sun}$)} & \multicolumn{1}{c}{(log $L/L_{\sun}$)} & \\ 
\hline

J06421+035 &GJ 3404A      	 &M3.5 V&3.090$\pm$0.602&2.486$\pm$0.346&1.820$\pm$0.099&3493$\pm$117&0.375$\pm$0.068&\textendash1.766$\pm$0.230& \\
J06548+332 &GJ 251        	 &M3.0 V&2.532$\pm$0.567&2.160$\pm$0.316&1.683$\pm$0.282&3366$\pm$114&0.332$\pm$0.068&\textendash1.936$\pm$0.229& \\
J06574+740 &2M J06572616+7405265 &M4.0 V&2.104$\pm$0.529&1.521$\pm$0.310&1.896$\pm$0.126&3287$\pm$124&0.324$\pm$0.068&\textendash1.933$\pm$0.228& \\
J07033+346 &GJ 3423       	 &M4.0 V&2.160$\pm$0.498&1.426$\pm$0.296&1.526$\pm$0.135&3278$\pm$121&0.311$\pm$0.069&\textendash1.964$\pm$0.234& \\
J07044+682 &GJ 258        	 &M3.0 V&2.192$\pm$0.636&2.015$\pm$0.381&2.459$\pm$0.224&3345$\pm$122&0.345$\pm$0.070&\textendash1.893$\pm$0.236& \\
J07274+052 &GJ 273        	 &M3.5 V&2.201$\pm$0.263&1.871$\pm$0.156&0.666$\pm$0.044&3256$\pm$113&0.282$\pm$0.068&\textendash1.924$\pm$0.228& \\
J07287-032 &GJ 1097       	 &M3.0 V&2.837$\pm$0.499&1.913$\pm$0.320&2.772$\pm$0.103&3493$\pm$111&0.418$\pm$0.067&\textendash1.597$\pm$0.223& \\
J07319+362N&GJ 277C       	 &M3.5 V&1.898$\pm$0.466&1.978$\pm$0.328&1.623$\pm$0.133&3228$\pm$109&0.285$\pm$0.067&\textendash2.158$\pm$0.225& \\
J07353+548 &GJ 3452       	 &M2.0 V&2.929$\pm$0.328&2.474$\pm$0.310&2.423$\pm$0.391&3489$\pm$117&0.387$\pm$0.069&\textendash1.758$\pm$0.232& \\
J07361-031 &GJ 282C       	 &M1.0 V&3.247$\pm$0.178&2.561$\pm$0.178&3.950$\pm$0.177&3662$\pm$107&0.497$\pm$0.067&\textendash1.382$\pm$0.224& \\
J07386-212 &GJ 3459       	 &M3.0 V&2.596$\pm$0.325&2.205$\pm$0.324&2.089$\pm$0.398&3400$\pm$115&0.352$\pm$0.068&\textendash1.872$\pm$0.230& \\
J07393+021 &GJ 281        	 &M0.0 V&4.329$\pm$0.235&3.435$\pm$0.211&4.928$\pm$0.069&3942$\pm$108&0.601$\pm$0.067&\textendash1.125$\pm$0.225& \\
J07472+503 &2M J07471385+5020386 &M4.0 V&2.193$\pm$0.269&1.664$\pm$0.316&1.500$\pm$0.158&3284$\pm$109&0.307$\pm$0.067&\textendash2.006$\pm$0.224& \\
J07558+833 &GJ 1101       	 &M4.5 V&1.313$\pm$0.297&1.025$\pm$0.367&1.437$\pm$0.216&3099$\pm$109&0.256$\pm$0.067&\textendash2.208$\pm$0.224& \\
J07582+413 &GJ 1105       	 &M3.5 V&2.550$\pm$0.305&1.257$\pm$0.358&2.172$\pm$0.172&3396$\pm$108&0.390$\pm$0.067&\textendash1.620$\pm$0.224& \\
J08119+087 &GJ 299        	 &M4.5 V&2.098$\pm$0.318&1.759$\pm$0.415&1.107$\pm$0.150&3246$\pm$109&0.285$\pm$0.067&\textendash2.082$\pm$0.223& \\
J08126-215 &GJ 300        	 &M4.0 V&1.118$\pm$0.265&1.443$\pm$0.368&1.849$\pm$0.419&3106$\pm$103&0.250$\pm$0.066&\textendash2.278$\pm$0.220& \\
J08161+013 &GJ 2066       	 &M2.0 V&3.431$\pm$0.183&2.750$\pm$0.196&2.449$\pm$0.187&3597$\pm$103&0.420$\pm$0.066&\textendash1.641$\pm$0.221& \\
J08293+039 &2M J08292191+0355092 &M2.5 V&2.991$\pm$0.166&2.144$\pm$0.220&3.318$\pm$0.144&3563$\pm$102&0.455$\pm$0.066&\textendash1.487$\pm$0.220& \\
J08315+730 &LP 035-219           &M4.0 V&1.772$\pm$0.328&1.496$\pm$0.431&1.693$\pm$0.115&3207$\pm$104&0.289$\pm$0.066&\textendash2.097$\pm$0.222& \\
J08358+680 &GJ 3506       	 &M2.5 V&2.756$\pm$0.246&1.946$\pm$0.344&1.964$\pm$0.180&3428$\pm$116&0.365$\pm$0.067&\textendash1.788$\pm$0.221& \\
J08402+314 &LSPM J0840+3127      &M3.5 V&2.538$\pm$0.246&2.091$\pm$0.381&1.317$\pm$0.199&3352$\pm$104&0.322$\pm$0.066&\textendash1.935$\pm$0.222& \\
J08413+594 &GJ 3512       	 &M5.5 V&0.566$\pm$0.423&1.164$\pm$0.639&0.107$\pm$0.204&2870$\pm$108&0.153$\pm$0.067&\textendash2.606$\pm$0.225& \\
J09005+465 &GJ 1119       	 &M4.5 V&2.158$\pm$0.282&1.456$\pm$0.404&1.292$\pm$0.263&3267$\pm$118&0.300$\pm$0.068&\textendash2.002$\pm$0.231& \\
J09028+680 &GJ 3526       	 &M4.0 V&2.139$\pm$0.252&1.835$\pm$0.412&1.549$\pm$0.163&3274$\pm$116&0.302$\pm$0.068&\textendash2.051$\pm$0.231& \\
J09133+688 &G 234-057       	 &M2.5 V&3.265$\pm$0.240&2.778$\pm$0.413&2.500$\pm$0.200&3564$\pm$111&0.409$\pm$0.067&\textendash1.690$\pm$0.226& \\
J09140+196 &LP 427-016           &M3.0 V&2.528$\pm$0.248&2.394$\pm$0.420&2.228$\pm$0.188&3395$\pm$114&0.349$\pm$0.068&\textendash1.905$\pm$0.229& \\
J09161+018 &RX J0916.1+0153      &M4.0 V&1.054$\pm$0.364&1.257$\pm$0.683&1.623$\pm$0.181&3072$\pm$117&0.238$\pm$0.069&\textendash2.325$\pm$0.236& \\
J09163-186 &GJ 3543       	 &M1.5 V&3.338$\pm$0.208&2.527$\pm$0.383&3.300$\pm$0.309&3629$\pm$105&0.462$\pm$0.067&\textendash1.493$\pm$0.222& \\
J09307+003 &GJ 1125       	 &M3.5 V&2.387$\pm$0.231&2.455$\pm$0.451&1.557$\pm$0.267&3328$\pm$117&0.314$\pm$0.068&\textendash2.036$\pm$0.227& \\
J09360-216 &GJ 357        	 &M2.5 V&2.474$\pm$0.204&2.294$\pm$0.382&2.394$\pm$0.435&3395$\pm$105&0.355$\pm$0.066&\textendash1.875$\pm$0.221& \\
J09411+132 &GJ 361        	 &M1.5 V&3.115$\pm$0.193&2.624$\pm$0.244&2.743$\pm$0.115&3547$\pm$103&0.412$\pm$0.066&\textendash1.677$\pm$0.221& \\
J09423+559 &GJ 363        	 &M3.5 V&2.190$\pm$0.211&1.883$\pm$0.397&1.510$\pm$0.093&3283$\pm$113&0.304$\pm$0.067&\textendash2.043$\pm$0.224& \\
J09425+700 &GJ 360        	 &M2.0 V&3.192$\pm$0.124&2.680$\pm$0.257&2.773$\pm$0.354&3564$\pm$102&0.418$\pm$0.066&\textendash1.659$\pm$0.220& \\
J09428+700 &GJ 362        	 &M3.0 V&2.985$\pm$0.142&2.206$\pm$0.244&2.171$\pm$0.113&3487$\pm$112&0.386$\pm$0.068&\textendash1.730$\pm$0.228& \\
J09439+269 &GJ 3564       	 &M3.5 V&2.680$\pm$0.302&2.355$\pm$0.343&1.439$\pm$0.117&3388$\pm$107&0.333$\pm$0.067&\textendash1.909$\pm$0.224& \\
J09447-182 &GJ 1129       	 &M4.0 V&1.465$\pm$0.469&1.767$\pm$0.521&1.180$\pm$0.184&3112$\pm$128&0.239$\pm$0.070&\textendash2.370$\pm$0.243& \\
J09468+760 &GJ 366        	 &M1.5 V&3.405$\pm$0.168&2.760$\pm$0.196&3.914$\pm$0.185&3688$\pm$106&0.496$\pm$0.067&\textendash1.400$\pm$0.225& \\
J09511-123 &GJ 369        	 &M0.5 V&3.728$\pm$0.155&2.907$\pm$0.193&3.390$\pm$0.279&3713$\pm$107&0.484$\pm$0.067&\textendash1.448$\pm$0.226& \\
J09561+627 &GJ 373        	 &M0.0 V&4.003$\pm$0.160&3.502$\pm$0.191&4.874$\pm$0.132&3878$\pm$106&0.570$\pm$0.067&\textendash1.221$\pm$0.224& \\
J10023+480 &GJ 378        	 &M1.0 V&3.949$\pm$0.251&2.947$\pm$0.296&4.633$\pm$0.182&3847$\pm$113&0.577$\pm$0.068&\textendash1.160$\pm$0.233& \\
J10122-037 &GJ 382        	 &M1.5 V&3.331$\pm$0.104&2.060$\pm$0.122&3.883$\pm$0.088&3672$\pm$104&0.531$\pm$0.066&\textendash1.227$\pm$0.221& \\
J10125+570 &LP 092-048           &M3.5 V&2.718$\pm$0.268&1.663$\pm$0.315&1.784$\pm$0.304&3411$\pm$114&0.360$\pm$0.068&\textendash1.768$\pm$0.231& \\
J10167-119 &GJ 386        	 &M3.0 V&2.709$\pm$0.252&2.540$\pm$0.305&3.597$\pm$0.242&3536$\pm$116&0.433$\pm$0.068&\textendash1.596$\pm$0.233& \\
J10251-102 &GJ 390        	 &M1.0 V&3.444$\pm$0.146&2.956$\pm$0.187&3.872$\pm$0.158&3692$\pm$108&0.488$\pm$0.067&\textendash1.441$\pm$0.225& \\
J10289+008 &GJ 393        	 &M2.0 V&3.210$\pm$0.149&2.745$\pm$0.136&2.784$\pm$0.077&3569$\pm$107&0.418$\pm$0.067&\textendash1.663$\pm$0.224& \\
J10350-094 &LP 670-017    	 &M3.0 V&2.596$\pm$0.288&2.112$\pm$0.330&2.547$\pm$0.491&3429$\pm$108&0.378$\pm$0.067&\textendash1.770$\pm$0.226& \\
J10396-069 &GJ 399        	 &M2.5 V&2.539$\pm$0.170&2.732$\pm$0.212&2.699$\pm$0.203&3429$\pm$104&0.364$\pm$0.066&\textendash1.874$\pm$0.222& \\
J10504+331 &GJ 3626       	 &M4.0 V&1.976$\pm$0.379&1.874$\pm$0.533&2.262$\pm$0.265&3289$\pm$126&0.323$\pm$0.069&\textendash1.976$\pm$0.235& \\
J10508+068 &GJ 402        	 &M4.0 V&1.425$\pm$0.259&1.466$\pm$0.395&0.577$\pm$0.190&3077$\pm$113&0.223$\pm$0.068&\textendash2.332$\pm$0.228& \\
J11000+228 &GJ 408        	 &M2.5 V&2.774$\pm$0.174&2.752$\pm$0.299&1.869$\pm$0.095&3427$\pm$102&0.350$\pm$0.066&\textendash1.908$\pm$0.220& \\
J11026+219 &GJ 410        	 &M2.0 V&3.354$\pm$0.119&2.544$\pm$0.237&4.553$\pm$0.139&3736$\pm$104&0.552$\pm$0.066&\textendash1.206$\pm$0.221& \\
J11033+359 &GJ 411        	 &M1.5 V&2.616$\pm$0.062&2.040$\pm$0.144&2.775$\pm$0.027&3450$\pm$103&0.395$\pm$0.066&\textendash1.695$\pm$0.220& \\
J11054+435 &GJ 412A       	 &M1.0 V&3.565$\pm$0.147&2.856$\pm$0.381&2.219$\pm$0.106&3615$\pm$107&0.420$\pm$0.067&\textendash1.630$\pm$0.224& \\
J11110+304 &GJ 414B       	 &M2.0 V&4.835$\pm$0.118&3.853$\pm$0.323&6.871$\pm$0.127&4207$\pm$104&0.759$\pm$0.066&\textendash0.726$\pm$0.222& \\
J11126+189 &GJ 3649       	 &M1.5 V&3.647$\pm$0.279&2.728$\pm$0.736&3.311$\pm$0.295&3691$\pm$112&0.479$\pm$0.067&\textendash1.450$\pm$0.226& \\
J11201-104 &LP 733-099    	 &M2.0 V&2.806$\pm$0.258&2.361$\pm$0.715&3.156$\pm$0.162&3515$\pm$107&0.420$\pm$0.067&\textendash1.631$\pm$0.225& \\
J11289+101 &GJ 3666       	 &M3.5 V&2.348$\pm$0.263&1.926$\pm$0.744&1.396$\pm$0.175&3313$\pm$113&0.311$\pm$0.068&\textendash1.994$\pm$0.229& \\
J11302+076 &K2-18    		 &M2.5 V&3.053$\pm$0.387&2.454$\pm$1.059&1.958$\pm$0.286&3491$\pm$105&0.377$\pm$0.066&\textendash1.767$\pm$0.222& \\
J11306-080 &LP 672-042           &M3.5 V&2.557$\pm$0.239&1.843$\pm$0.676&1.794$\pm$0.112&3376$\pm$122&0.344$\pm$0.069&\textendash1.863$\pm$0.237& \\
J11417+427 &GJ 1148       	 &M4.0 V&1.468$\pm$0.263&1.954$\pm$0.797&1.176$\pm$0.244&3112$\pm$107&0.238$\pm$0.067&\textendash2.392$\pm$0.224& \\
J11421+267 &GJ 436        	 &M2.5 V&2.593$\pm$0.197&2.072$\pm$0.578&2.104$\pm$0.079&3401$\pm$118&0.356$\pm$0.068&\textendash1.847$\pm$0.234& \\
J11467-140 &GJ 443        	 &M3.0 V&3.012$\pm$0.237&1.784$\pm$0.715&3.592$\pm$0.288&3589$\pm$107&0.505$\pm$0.067&\textendash1.286$\pm$0.228& \\
J11476+002 &GJ 3685A      	 &M4.0 V&0.968$\pm$0.284&1.867$\pm$0.917&1.197$\pm$0.433&3015$\pm$116&0.201$\pm$0.068&\textendash2.603$\pm$0.231& \\

\hline
\end{tabular}
     \raggedright \hspace*{14.5 cm} (continued on next page ..) \\
\end{table*}

\begin{table*}
 \centering
 \contcaption{from the previous page.}
 \begin{tabular}{lcccccccccccc}
  \hline \hline

\hline \multicolumn{1}{c}{Karmn$^{(a)}$} & \multicolumn{1}{c}{Name} & \multicolumn{1}{c}{Sp.type$^{(b)}$} & \multicolumn{1}{c}{Ca I EW (${\AA}$)} & \multicolumn{1}{c}{Ca I EW (${\AA}$)} & \multicolumn{1}{c}{Mg I EW (${\AA}$)} & \multicolumn{1}{c}{$T_{eff}^{(c)}$} & \multicolumn{1}{c}{Radius$^{(c)}$} & \multicolumn{1}{c}{Luminosity$^{(c)}$} & \\ 

\multicolumn{1}{c}{} & \multicolumn{1}{c}{} & \multicolumn{1}{c}{} & \multicolumn{1}{c}{(0.854 $\mu$m)} & \multicolumn{1}{c}{(0.866 $\mu$m)} & \multicolumn{1}{c}{(1.57 $\mu$m)} & \multicolumn{1}{c}{(K)} & \multicolumn{1}{c}{($R_{\sun}$)} & \multicolumn{1}{c}{(log $L/L_{\sun}$)} & \\ 
\hline

J11476+786 &GJ 445        	 &M3.5 V&2.199$\pm$0.254&1.614$\pm$0.742&1.755$\pm$0.136&3298$\pm$111&0.321$\pm$0.067&\textendash1.952$\pm$0.226& \\
J11477+008 &GJ 447        	 &M4.0 V&3.043$\pm$0.252&2.786$\pm$0.734&2.907$\pm$0.346&3543$\pm$109&0.411$\pm$0.067&\textendash1.698$\pm$0.225& \\
J11509+483 &GJ 1151       	 &M4.5 V&1.344$\pm$0.265&1.295$\pm$0.783&0.654$\pm$0.255&3062$\pm$107&0.220$\pm$0.067&\textendash2.384$\pm$0.224& \\
J11511+352 &GJ 450        	 &M1.5 V&3.492$\pm$0.197&2.895$\pm$0.577&2.756$\pm$0.204&3626$\pm$105&0.435$\pm$0.067&\textendash1.607$\pm$0.222& \\
J12054+695 &GJ 3704       	 &M4.0 V&1.132$\pm$0.365&1.997$\pm$1.027&0.909$\pm$0.338&3027$\pm$108&0.205$\pm$0.067&\textendash2.580$\pm$0.224& \\
J12100-150 &GJ 3707       	 &M3.5 V&2.420$\pm$0.300&1.933$\pm$0.806&1.903$\pm$0.217&3353$\pm$109&0.337$\pm$0.067&\textendash1.913$\pm$0.224& \\
J12111-199 &GJ 3708A      	 &M3.0 V&2.964$\pm$0.195&2.479$\pm$0.520&2.589$\pm$0.207&3506$\pm$105&0.397$\pm$0.067&\textendash1.722$\pm$0.222& \\
J12123+544S&GJ 458A       	 &M0.0 V&3.970$\pm$0.159&2.856$\pm$0.429&5.066$\pm$0.135&3889$\pm$113&0.618$\pm$0.069&\textendash1.035$\pm$0.227& \\
J12156+526 &StKM 2-809    	 &M4.0 V&2.146$\pm$0.268&1.633$\pm$0.696&1.830$\pm$0.194&3292$\pm$111&0.320$\pm$0.067&\textendash1.961$\pm$0.224& \\
J12189+111 &GJ 1156       	 &M5.0 V&0.849$\pm$0.393&1.217$\pm$0.992&0.025$\pm$0.228&2935$\pm$108&0.174$\pm$0.067&\textendash0.865$\pm$0.224& \\
J12230+640 &GJ 463        	 &M3.0 V&2.959$\pm$0.415&2.590$\pm$1.084&2.746$\pm$0.202&3515$\pm$114&0.401$\pm$0.069&\textendash1.717$\pm$0.228& \\
J12248-182 &GJ 465        	 &M2.0 V&3.457$\pm$0.290&2.746$\pm$0.752&1.718$\pm$0.321&3571$\pm$104&0.397$\pm$0.067&\textendash1.659$\pm$0.222& \\
J12312+086 &GJ 471        	 &M0.5 V&4.129$\pm$0.291&2.588$\pm$0.731&4.746$\pm$0.146&3890$\pm$123&0.624$\pm$0.071&\textendash0.996$\pm$0.231& \\
J12350+098 &GJ 476        	 &M2.5 V&3.434$\pm$0.261&2.579$\pm$0.624&3.084$\pm$0.226&3633$\pm$106&0.455$\pm$0.067&\textendash1.519$\pm$0.222& \\
J12373-208 &LP 795-038    	 &M4.0 V&2.050$\pm$0.458&1.658$\pm$1.152&1.639$\pm$0.285&3261$\pm$112&0.303$\pm$0.068&\textendash2.039$\pm$0.227& \\
J12388+116 &GJ 480        	 &M3.0 V&2.296$\pm$0.297&2.056$\pm$0.742&2.567$\pm$0.136&3373$\pm$105&0.358$\pm$0.067&\textendash1.844$\pm$0.222& \\
J12428+418 &G 123-055    	 &M4.0 V&1.304$\pm$0.351&1.941$\pm$0.937&0.396$\pm$0.337&3045$\pm$106&0.211$\pm$0.067&\textendash2.339$\pm$0.223& \\
J12479+097 &GJ 486        	 &M3.5 V&2.388$\pm$0.348&1.633$\pm$0.791&1.564$\pm$0.312&3329$\pm$105&0.326$\pm$0.067&\textendash1.915$\pm$0.223& \\
J13005+056 &GJ 493.1      	 &M4.5 V&1.088$\pm$0.384&0.709$\pm$0.774&0.965$\pm$0.448&3021$\pm$117&0.223$\pm$0.068&\textendash2.349$\pm$0.227& \\
J13102+477 &G 177-025            &M5.0 V&1.261$\pm$0.652&1.233$\pm$1.300&0.717$\pm$0.214&3045$\pm$114&0.215$\pm$0.068&\textendash2.430$\pm$0.228& \\
J13196+333 &GJ 507.1      	 &M1.5 V&3.532$\pm$0.303&3.223$\pm$0.635&3.857$\pm$0.221&3707$\pm$105&0.484$\pm$0.067&\textendash1.473$\pm$0.223& \\
J13209+342 &GJ 508.2      	 &M1.0 V&3.507$\pm$0.223&2.900$\pm$0.393&4.153$\pm$0.134&3727$\pm$104&0.513$\pm$0.066&\textendash1.355$\pm$0.221& \\
J13229+244 &GJ 3779       	 &M4.0 V&2.081$\pm$0.373&2.146$\pm$0.652&0.822$\pm$0.079&3233$\pm$106&0.275$\pm$0.067&\textendash2.081$\pm$0.222& \\
J13283-023W&GJ 512A       	 &M3.0 V&2.697$\pm$0.226&1.797$\pm$0.378&2.983$\pm$0.159&3481$\pm$111&0.429$\pm$0.066&\textendash1.548$\pm$0.221& \\
J13293+114 &GJ 513        	 &M3.5 V&2.948$\pm$0.427&2.424$\pm$0.637&2.249$\pm$0.443&3483$\pm$107&0.382$\pm$0.067&\textendash1.769$\pm$0.223& \\
J13299+102 &GJ 514        	 &M0.5 V&3.537$\pm$0.167&3.146$\pm$0.274&3.341$\pm$0.111&3671$\pm$103&0.460$\pm$0.066&\textendash1.548$\pm$0.220& \\
J13427+332 &GJ 3801       	 &M3.5 V&1.948$\pm$0.524&1.573$\pm$0.638&1.186$\pm$0.202&3216$\pm$107&0.278$\pm$0.067&\textendash2.123$\pm$0.224& \\
J13450+176 &GJ 525        	 &M1.0 V&3.694$\pm$0.295&3.163$\pm$0.342&3.359$\pm$0.138&3704$\pm$103&0.472$\pm$0.066&\textendash1.506$\pm$0.221& \\
J13457+148 &GJ 526        	 &M1.5 V&3.438$\pm$0.266&2.114$\pm$0.280&3.524$\pm$0.090&3664$\pm$105&0.506$\pm$0.067&\textendash1.309$\pm$0.222& \\
J13458-179 &GJ 3804       	 &M3.5 V&2.890$\pm$0.658&2.275$\pm$0.667&1.558$\pm$0.113&3439$\pm$111&0.354$\pm$0.068&\textendash1.818$\pm$0.228& \\
J13536+776 &RX J1353.6+7737      &M4.0 V&1.514$\pm$0.661&1.320$\pm$0.665&0.630$\pm$0.358&3099$\pm$113&0.232$\pm$0.068&\textendash2.278$\pm$0.229& \\
J13582+125 &GJ 3817       	 &M3.0 V&1.880$\pm$0.599&2.310$\pm$0.619&0.707$\pm$0.161&3184$\pm$114&0.258$\pm$0.068&\textendash2.162$\pm$0.230& \\
J14010-026 &GJ 536        	 &M1.0 V&3.357$\pm$0.349&2.766$\pm$0.343&3.848$\pm$0.143&3674$\pm$105&0.488$\pm$0.066&\textendash1.427$\pm$0.221& \\
J14082+805 &GJ 540        	 &M1.0 V&3.673$\pm$0.504&3.074$\pm$0.464&4.530$\pm$0.437&3789$\pm$114&0.542$\pm$0.070&\textendash1.275$\pm$0.235& \\
J14152+450 &GJ 3836       	 &M3.0 V&2.183$\pm$0.662&2.521$\pm$0.595&2.041$\pm$0.248&3313$\pm$118&0.314$\pm$0.070&\textendash2.068$\pm$0.234& \\
J14173+454 &RX J1417.3+4525      &M5.0 V&1.215$\pm$1.670&1.532$\pm$1.172&0.031$\pm$0.351&3020$\pm$111&0.202$\pm$0.067&\textendash0.114$\pm$0.227& \\
J14251+518 &GJ 549B       	 &M2.5 V&3.159$\pm$0.475&2.024$\pm$0.380&2.314$\pm$0.164&3532$\pm$108&0.410$\pm$0.067&\textendash1.617$\pm$0.224& \\
J14257+236E&GJ 548B       	 &M0.5 V&4.243$\pm$0.416&3.348$\pm$0.327&4.938$\pm$0.175&3927$\pm$106&0.600$\pm$0.067&\textendash1.124$\pm$0.222& \\
J14257+236W&GJ 548A       	 &M0.0 V&4.295$\pm$0.368&3.526$\pm$0.297&5.366$\pm$0.150&3973$\pm$104&0.624$\pm$0.066&\textendash1.066$\pm$0.221& \\
J14294+155 &GJ 552        	 &M2.0 V&3.492$\pm$0.659&2.738$\pm$0.526&3.046$\pm$0.306&3643$\pm$108&0.453$\pm$0.067&\textendash1.539$\pm$0.226& \\
J14307-086 &GJ 553        	 &M0.5 V&4.218$\pm$0.491&3.229$\pm$0.361&6.351$\pm$0.135&4056$\pm$103&0.718$\pm$0.066&\textendash0.791$\pm$0.221& \\
J14310-122 &GJ 553.1      	 &M3.5 V&2.066$\pm$0.771&1.911$\pm$0.531&1.630$\pm$0.181&3263$\pm$104&0.299$\pm$0.066&\textendash2.081$\pm$0.221& \\
J14342-125 &GJ 555        	 &M4.0 V&1.854$\pm$0.785&1.508$\pm$0.486&2.160$\pm$0.201&3258$\pm$108&0.322$\pm$0.066&\textendash1.945$\pm$0.220& \\
J14524+123 &GJ 3871       	 &M2.0 V&2.975$\pm$0.707&2.322$\pm$0.467&3.215$\pm$0.281&3552$\pm$104&0.438$\pm$0.066&\textendash1.562$\pm$0.220& \\
J14544+355 &GJ 3873       	 &M3.5 V&2.385$\pm$0.858&2.336$\pm$0.576&1.449$\pm$0.192&3323$\pm$117&0.312$\pm$0.068&\textendash2.027$\pm$0.231& \\
J15013+055 &GJ 3885       	 &M3.0 V&2.014$\pm$0.559&2.462$\pm$0.401&1.710$\pm$0.290&3257$\pm$110&0.290$\pm$0.067&\textendash2.167$\pm$0.226& \\
J15095+031 &GJ 3892       	 &M3.0 V&3.016$\pm$0.703&2.323$\pm$0.467&2.929$\pm$0.132&3539$\pm$114&0.424$\pm$0.067&\textendash1.613$\pm$0.226& \\
J15194-077 &GJ 581        	 &M3.0 V&2.424$\pm$0.461&2.162$\pm$0.308&1.722$\pm$0.293&3344$\pm$108&0.325$\pm$0.067&\textendash1.974$\pm$0.224& \\
J15218+209 &GJ 9520       	 &M1.5 V&2.157$\pm$0.420&1.992$\pm$0.286&3.664$\pm$0.373&3460$\pm$122&0.429$\pm$0.072&\textendash1.562$\pm$0.231& \\
J15369-141 &GJ 592        	 &M4.0 V&2.345$\pm$0.985&2.076$\pm$0.635&2.072$\pm$0.069&3347$\pm$109&0.336$\pm$0.067&\textendash1.937$\pm$0.225& \\
J15499+796 &LP 022-420           &M5.0 V&1.152$\pm$1.329&0.509$\pm$0.823&0.956$\pm$0.258&3034$\pm$113&0.239$\pm$0.068&\textendash2.211$\pm$0.227& \\
J15598-082 &GJ 606        	 &M1.0 V&3.287$\pm$0.432&2.669$\pm$0.292&3.950$\pm$0.348&3669$\pm$104&0.494$\pm$0.067&\textendash1.399$\pm$0.222& \\
J16028+205 &GJ 609        	 &M4.0 V&2.514$\pm$0.691&1.411$\pm$0.434&1.177$\pm$0.303&3341$\pm$106&0.323$\pm$0.067&\textendash1.840$\pm$0.224& \\
J16092+093 &G 137-084            &M3.0 V&2.474$\pm$0.648&2.099$\pm$0.428&2.186$\pm$0.152&3381$\pm$105&0.350$\pm$0.066&\textendash1.877$\pm$0.222& \\
J16167+672N&GJ 617B       	 &M3.0 V&2.908$\pm$0.375&2.355$\pm$0.242&2.928$\pm$0.113&3518$\pm$104&0.414$\pm$0.066&\textendash1.651$\pm$0.221& \\
J16167+672S&GJ 617A       	 &M0.0 V&4.421$\pm$0.590&3.784$\pm$0.412&6.344$\pm$0.346&4087$\pm$106&0.691$\pm$0.068&\textendash0.898$\pm$0.225& \\
J16254+543 &GJ 625        	 &M1.5 V&3.010$\pm$0.309&2.018$\pm$0.205&1.875$\pm$0.184&3478$\pm$103&0.378$\pm$0.066&\textendash1.722$\pm$0.221& \\
J16303-126 &GJ 628        	 &M3.5 V&2.548$\pm$0.303&1.753$\pm$0.182&1.528$\pm$0.212&3362$\pm$104&0.334$\pm$0.066&\textendash1.878$\pm$0.222& \\
J16327+126 &GJ 1203       	 &M3.0 V&3.019$\pm$0.574&2.529$\pm$0.404&2.463$\pm$0.109&3510$\pm$112&0.394$\pm$0.068&\textendash1.734$\pm$0.227& \\
J16462+164 &GJ 3972       	 &M2.5 V&2.439$\pm$0.352&2.417$\pm$0.236&2.232$\pm$0.432&3377$\pm$107&0.342$\pm$0.067&\textendash1.937$\pm$0.224& \\
J16554-083N&GJ 643        	 &M3.5 V&2.035$\pm$0.792&1.373$\pm$0.515&1.356$\pm$0.145&3243$\pm$106&0.295$\pm$0.067&\textendash2.030$\pm$0.223& \\
J16570-043 &GJ 1207       	 &M3.5 V&1.724$\pm$0.516&1.311$\pm$0.402&1.587$\pm$0.174&3190$\pm$125&0.285$\pm$0.070&\textendash2.093$\pm$0.237& \\
J16581+257 &GJ 649        	 &M1.0 V&3.399$\pm$0.263&2.753$\pm$0.211&3.523$\pm$0.035&3657$\pm$107&0.472$\pm$0.067&\textendash1.481$\pm$0.224& \\
J17033+514 &GJ 3988       	 &M4.5 V&0.919$\pm$0.561&1.364$\pm$0.411&0.563$\pm$0.161&2964$\pm$119&0.185$\pm$0.069&\textendash2.648$\pm$0.233& \\
J17052-050 &GJ 654        	 &M1.5 V&3.631$\pm$0.338&2.992$\pm$0.268&2.895$\pm$0.105&3663$\pm$110&0.450$\pm$0.067&\textendash1.563$\pm$0.226& \\

\hline
\end{tabular}
     \raggedright \hspace*{14.5 cm} (continued on next page ..) \\
\end{table*}

\begin{table*}
 \centering
 \contcaption{from the previous page.}
 \begin{tabular}{lcccccccccccc}
  \hline \hline

\hline \multicolumn{1}{c}{Karmn$^{(a)}$} & \multicolumn{1}{c}{Name} & \multicolumn{1}{c}{Sp.type$^{(b)}$} & \multicolumn{1}{c}{Ca I EW (${\AA}$)} & \multicolumn{1}{c}{Ca I EW (${\AA}$)} & \multicolumn{1}{c}{Mg I EW (${\AA}$)} & \multicolumn{1}{c}{$T_{eff}^{(c)}$} & \multicolumn{1}{c}{Radius$^{(c)}$} & \multicolumn{1}{c}{Luminosity$^{(c)}$} & \\ 

\multicolumn{1}{c}{} & \multicolumn{1}{c}{} & \multicolumn{1}{c}{} & \multicolumn{1}{c}{(0.854 $\mu$m)} & \multicolumn{1}{c}{(0.866 $\mu$m)} & \multicolumn{1}{c}{(1.57 $\mu$m)} & \multicolumn{1}{c}{(K)} & \multicolumn{1}{c}{($R_{\sun}$)} & \multicolumn{1}{c}{(log $L/L_{\sun}$)} & \\ 
\hline

J17071+215 &GJ 655        	 &M3.0 V&2.326$\pm$0.563&2.224$\pm$0.428&2.536$\pm$0.180&3376$\pm$120&0.353$\pm$0.069&\textendash1.879$\pm$0.234& \\
J17115+384 &GJ 3992       	 &M3.5 V&2.622$\pm$0.361&1.751$\pm$0.248&2.164$\pm$0.093&3410$\pm$114&0.370$\pm$0.068&\textendash1.756$\pm$0.225& \\
J17166+080 &GJ 2128       	 &M2.0 V&3.186$\pm$0.557&1.897$\pm$0.396&2.543$\pm$0.233&3550$\pm$112&0.430$\pm$0.068&\textendash1.538$\pm$0.228& \\
J17198+417 &GJ 671        	 &M2.5 V&2.872$\pm$0.574&2.012$\pm$0.422&2.188$\pm$0.803&3464$\pm$119&0.383$\pm$0.070&\textendash1.728$\pm$0.233& \\
J17303+055 &GJ 678.1 A    	 &M0.0 V&3.419$\pm$0.308&2.306$\pm$0.214&4.081$\pm$0.105&3704$\pm$112&0.383$\pm$0.068&\textendash1.236$\pm$0.227& \\
J17355+616 &GJ 685        	 &M0.5 V&3.573$\pm$0.298&2.964$\pm$0.203&4.923$\pm$0.118&3808$\pm$107&0.570$\pm$0.068&\textendash1.184$\pm$0.225& \\
J17378+185 &GJ 686        	 &M1.0 V&3.570$\pm$0.301&2.776$\pm$0.175&3.228$\pm$0.107&3670$\pm$105&0.467$\pm$0.067&\textendash1.494$\pm$0.223& \\
J17542+073 &GJ 1222       	 &M4.0 V&1.955$\pm$0.732&1.765$\pm$0.467&1.398$\pm$0.300&3227$\pm$121&0.284$\pm$0.069&\textendash2.132$\pm$0.235& \\
J17578+046 &GJ 699        	 &M3.5 V&0.160$\pm$0.290&0.642$\pm$0.159&1.473$\pm$0.085&3256$\pm$111&0.200$\pm$0.066&\textendash2.455$\pm$0.220& \\
J17578+465 &GJ 4040       	 &M2.5 V&2.747$\pm$0.689&2.085$\pm$0.396&2.329$\pm$0.089&3446$\pm$117&0.378$\pm$0.069&\textendash1.760$\pm$0.235& \\
J18022+642 &LP 071-082     	 &M5.0 V&1.283$\pm$0.960&0.840$\pm$0.547&0.485$\pm$0.170&3043$\pm$112&0.214$\pm$0.068&\textendash2.320$\pm$0.228& \\
J18051-030 &GJ 701        	 &M1.0 V&3.581$\pm$0.221&3.037$\pm$0.147&2.777$\pm$0.143&3646$\pm$106&0.440$\pm$0.067&\textendash1.599$\pm$0.223& \\
J18174+483 &TYC 3529-1437-1      &M2.0 V&2.834$\pm$0.282&2.133$\pm$0.182&3.654$\pm$0.127&3563$\pm$112&0.469$\pm$0.068&\textendash1.438$\pm$0.226& \\
J18180+387E&GJ 4048A      	 &M3.0 V&2.531$\pm$0.449&2.490$\pm$0.337&1.026$\pm$0.167&3340$\pm$112&0.312$\pm$0.068&\textendash1.934$\pm$0.228& \\
J18189+661 &GJ 4053       	 &M4.5 V&1.463$\pm$0.491&1.501$\pm$0.339&0.283$\pm$0.246&3080$\pm$115&0.222$\pm$0.068&\textendash2.036$\pm$0.230& \\
J18221+063 &GJ 712        	 &M4.0 V&2.849$\pm$0.484&2.448$\pm$0.337&1.676$\pm$0.160&3434$\pm$114&0.352$\pm$0.068&\textendash1.854$\pm$0.230& \\
J18224+620 &GJ 1227       	 &M4.0 V&1.528$\pm$0.455&1.056$\pm$0.335&0.767$\pm$0.180&3107$\pm$112&0.239$\pm$0.068&\textendash2.252$\pm$0.228& \\
J18319+406 &GJ 4062       	 &M3.5 V&2.766$\pm$0.428&2.431$\pm$0.372&1.582$\pm$0.351&3412$\pm$115&0.343$\pm$0.069&\textendash1.885$\pm$0.231& \\
J18346+401 &GJ 4063       	 &M3.5 V&1.823$\pm$0.368&1.303$\pm$0.285&2.347$\pm$0.075&3269$\pm$118&0.346$\pm$0.068&\textendash1.821$\pm$0.227& \\
J18353+457 &GJ 720A       	 &M0.5 V&4.069$\pm$0.197&3.264$\pm$0.163&5.334$\pm$0.108&3931$\pm$121&0.620$\pm$0.072&\textendash1.060$\pm$0.236& \\
J18363+136 &GJ 4065       	 &M4.0 V&2.093$\pm$0.333&2.228$\pm$0.269&1.539$\pm$0.467&3264$\pm$120&0.293$\pm$0.069&\textendash2.127$\pm$0.231& \\
J18409-133 &GJ 724        	 &M1.0 V&3.522$\pm$0.206&2.960$\pm$0.173&4.598$\pm$0.077&3768$\pm$103&0.542$\pm$0.066&\textendash1.268$\pm$0.221& \\
J18419+318 &GJ 4070       	 &M3.0 V&2.350$\pm$0.370&2.210$\pm$0.311&2.026$\pm$0.133&3346$\pm$103&0.331$\pm$0.066&\textendash1.967$\pm$0.221& \\
J18480-145 &GJ 4077       	 &M2.5 V&2.445$\pm$0.388&2.109$\pm$0.320&2.254$\pm$0.276&3379$\pm$107&0.351$\pm$0.067&\textendash1.876$\pm$0.223& \\
J18498-238 &GJ 729        	 &M3.5 V&1.898$\pm$0.228&2.323$\pm$0.200&1.225$\pm$0.281&3207$\pm$107&0.269$\pm$0.067&\textendash2.235$\pm$0.224& \\
J18580+059 &GJ 740        	 &M0.5 V&3.757$\pm$0.213&2.903$\pm$0.172&4.842$\pm$0.331&3831$\pm$112&0.581$\pm$0.068&\textendash1.146$\pm$0.228& \\
J19070+208 &GJ 745A       	 &M2.0 V&3.224$\pm$0.280&2.773$\pm$0.238&1.594$\pm$0.264&3514$\pm$116&0.375$\pm$0.069&\textendash1.742$\pm$0.232& \\
J19072+208 &GJ 745B       	 &M2.0 V&3.083$\pm$0.412&2.758$\pm$0.387&1.660$\pm$0.385&3485$\pm$125&0.367$\pm$0.070&\textendash1.798$\pm$0.239& \\
J19084+322 &GJ 4098       	 &M3.0 V&2.035$\pm$0.403&2.409$\pm$0.353&1.690$\pm$0.216&3260$\pm$120&0.292$\pm$0.070&\textendash2.156$\pm$0.235& \\
J19098+176 &GJ 1232       	 &M4.5 V&1.330$\pm$0.494&1.882$\pm$0.416&0.331$\pm$0.344&3050$\pm$105&0.212$\pm$0.067&\textendash2.245$\pm$0.223& \\
J19169+051N&GJ 752A       	 &M2.5 V&3.094$\pm$0.203&2.806$\pm$0.174&2.667$\pm$0.343&3538$\pm$107&0.403$\pm$0.067&\textendash1.724$\pm$0.224& \\
J19216+208 &GJ 1235       	 &M4.5 V&1.082$\pm$0.501&1.110$\pm$0.367&0.687$\pm$0.162&3005$\pm$113&0.202$\pm$0.068&\textendash2.517$\pm$0.226& \\
J19251+283 &GJ 4109       	 &M3.0 V&2.087$\pm$0.434&1.873$\pm$0.361&1.811$\pm$0.255&3278$\pm$125&0.309$\pm$0.071&\textendash2.037$\pm$0.238& \\
J19346+045 &GJ 763        	 &M0.0 V&4.063$\pm$0.217&3.351$\pm$0.189&5.803$\pm$0.147&3975$\pm$108&0.650$\pm$0.068&\textendash0.980$\pm$0.225& \\
J19511+464 &GJ 1243       	 &M4.0 V&1.624$\pm$0.412&1.532$\pm$0.367&1.079$\pm$0.273&3141$\pm$113&0.251$\pm$0.068&\textendash2.273$\pm$0.228& \\
J20260+585 &GJ 1253       	 &M5.0 V&1.511$\pm$0.781&1.421$\pm$0.698&1.091$\pm$0.263&3117$\pm$106&0.244$\pm$0.067&\textendash2.307$\pm$0.226& \\
J20305+654 &GJ 793        	 &M2.5 V&2.417$\pm$0.263&2.413$\pm$0.236&1.874$\pm$0.112&3351$\pm$109&0.327$\pm$0.067&\textendash1.994$\pm$0.224& \\
J20336+617 &GJ 1254       	 &M4.0 V&1.373$\pm$0.434&1.403$\pm$0.398&2.357$\pm$0.271&3200$\pm$106&0.306$\pm$0.067&\textendash2.001$\pm$0.222& \\
J20405+154 &GJ 1256       	 &M4.5 V&0.962$\pm$0.444&0.602$\pm$0.381&0.905$\pm$0.103&2992$\pm$111&0.214$\pm$0.067&\textendash2.385$\pm$0.226& \\
J20450+444 &GJ 806        	 &M1.5 V&3.245$\pm$0.256&2.437$\pm$0.259&2.856$\pm$0.061&3580$\pm$112&0.433$\pm$0.067&\textendash1.585$\pm$0.224& \\
J20525-169 &LP 816-060           &M4.0 V&1.139$\pm$0.423&1.361$\pm$0.457&1.479$\pm$0.183&3071$\pm$122&0.234$\pm$0.068&\textendash2.368$\pm$0.229& \\
J20533+621 &GJ 809A       	 &M0.5 V&3.887$\pm$0.187&3.028$\pm$0.188&4.644$\pm$0.126&3837$\pm$111&0.569$\pm$0.067&\textendash1.192$\pm$0.223& \\
J20567-104 &GJ 811.1      	 &M2.5 V&2.892$\pm$0.650&2.601$\pm$0.673&2.700$\pm$0.219&3499$\pm$110&0.394$\pm$0.067&\textendash1.746$\pm$0.226& \\
J21019-063 &GJ 816        	 &M2.5 V&2.334$\pm$0.401&2.478$\pm$0.435&2.603$\pm$0.175&3383$\pm$105&0.350$\pm$0.066&\textendash1.911$\pm$0.222& \\
J21152+257 &GJ 4184       	 &M3.0 V&2.957$\pm$0.451&2.571$\pm$0.469&3.586$\pm$0.237&3579$\pm$115&0.450$\pm$0.068&\textendash1.542$\pm$0.226& \\
J21164+025 &LSPM J2116+0234      &M3.0 V&2.989$\pm$0.425&2.296$\pm$0.428&2.365$\pm$0.326&3498$\pm$111&0.393$\pm$0.067&\textendash1.717$\pm$0.223& \\
J21348+515 &GJ 4205       	 &M3.0 V&2.506$\pm$0.400&2.246$\pm$0.421&2.660$\pm$0.309&3420$\pm$112&0.373$\pm$0.067&\textendash1.802$\pm$0.225& \\
J21463+382 &LSPM J2146+3813	 &M4.0 V&2.390$\pm$0.467&1.922$\pm$0.453&1.151$\pm$0.137&3313$\pm$111&0.307$\pm$0.067&\textendash1.970$\pm$0.225& \\
J21466+668 &G 264-012            &M4.0 V&2.065$\pm$0.413&2.138$\pm$0.444&1.636$\pm$0.043&3264$\pm$109&0.295$\pm$0.067&\textendash2.115$\pm$0.225& \\                                                                                                                                 
J22012+283 &GJ 4247       	 &M4.0 V&1.357$\pm$0.419&1.441$\pm$0.429&1.270$\pm$0.169&3095$\pm$113&0.239$\pm$0.068&\textendash2.349$\pm$0.228& \\
J22020-194 &GJ 843        	 &M3.5 V&2.013$\pm$0.387&1.668$\pm$0.478&2.613$\pm$0.185&3326$\pm$119&0.357$\pm$0.068&\textendash1.811$\pm$0.228& \\
J22021+014 &GJ 846        	 &M0.5 V&3.772$\pm$0.193&2.864$\pm$0.236&4.570$\pm$0.145&3810$\pm$104&0.564$\pm$0.067&\textendash1.194$\pm$0.222& \\
J22057+656 &GJ 4258       	 &M1.5 V&3.047$\pm$0.286&2.710$\pm$0.388&3.310$\pm$0.155&3573$\pm$109&0.434$\pm$0.067&\textendash1.608$\pm$0.225& \\
J22096-046 &GJ 849        	 &M3.5 V&2.742$\pm$0.180&1.906$\pm$0.296&3.303$\pm$0.115&3516$\pm$105&0.450$\pm$0.066&\textendash1.484$\pm$0.222& \\
J22114+409 &1RXS J221124.3+410000&M5.5 V&0.638$\pm$0.555&0.494$\pm$0.985&0.602$\pm$0.103&2907$\pm$105&0.175$\pm$0.066&\textendash2.644$\pm$0.223& \\
J22115+184 &GJ 851        	 &M2.0 V&2.758$\pm$0.149&2.374$\pm$0.294&3.867$\pm$0.179&3571$\pm$102&0.464$\pm$0.066&\textendash1.473$\pm$0.220& \\
J22125+085 &GJ 9773       	 &M3.0 V&2.971$\pm$0.195&2.524$\pm$0.375&2.071$\pm$0.167&3479$\pm$108&0.374$\pm$0.067&\textendash1.796$\pm$0.224& \\
J22137-176 &GJ 1265       	 &M4.5 V&1.150$\pm$0.392&1.535$\pm$0.772&1.129$\pm$0.121&3045$\pm$114&0.216$\pm$0.068&\textendash2.487$\pm$0.229& \\
J22252+594 &GJ 4276       	 &M4.0 V&2.088$\pm$0.255&1.307$\pm$0.542&2.182$\pm$0.160&3304$\pm$126&0.352$\pm$0.067&\textendash1.790$\pm$0.224& \\
J22298+414 &GJ 1270       	 &M4.0 V&1.679$\pm$0.277&2.158$\pm$0.566&0.588$\pm$0.117&3135$\pm$111&0.241$\pm$0.067&\textendash2.219$\pm$0.227& \\
J22330+093 &GJ 863        	 &M1.0 V&3.074$\pm$0.277&2.832$\pm$0.541&3.242$\pm$0.304&3573$\pm$112&0.429$\pm$0.068&\textendash1.637$\pm$0.228& \\
J22468+443 &GJ 873        	 &M3.5 V&1.312$\pm$0.257&2.024$\pm$0.518&1.281$\pm$0.191&3087$\pm$108&0.228$\pm$0.067&\textendash2.455$\pm$0.225& \\
J22503-070 &GJ 875        	 &M0.5 V&4.021$\pm$0.151&3.270$\pm$0.348&4.309$\pm$0.223&3835$\pm$112&0.546$\pm$0.068&\textendash1.280$\pm$0.229& \\
J22518+317 &GJ 875.1      	 &M3.0 V&0.794$\pm$0.237&0.893$\pm$0.477&1.606$\pm$0.300&3038$\pm$109&0.237$\pm$0.067&\textendash2.288$\pm$0.225& \\

\hline
\end{tabular}
     \raggedright \hspace*{14.5 cm} (continued on next page ..) \\
\end{table*}

\begin{table*}
 \centering
 \contcaption{from the previous page.}
 \begin{tabular}{lcccccccccccc}
  \hline \hline

\hline \multicolumn{1}{c}{Karmn$^{(a)}$} & \multicolumn{1}{c}{Name} & \multicolumn{1}{c}{Sp.type$^{(b)}$} & \multicolumn{1}{c}{Ca I EW (${\AA}$)} & \multicolumn{1}{c}{Ca I EW (${\AA}$)} & \multicolumn{1}{c}{Mg I EW (${\AA}$)} & \multicolumn{1}{c}{$T_{eff}^{(c)}$} & \multicolumn{1}{c}{Radius$^{(c)}$} & \multicolumn{1}{c}{Luminosity$^{(c)}$} & \\ 

\multicolumn{1}{c}{} & \multicolumn{1}{c}{} & \multicolumn{1}{c}{} & \multicolumn{1}{c}{(0.854 $\mu$m)} & \multicolumn{1}{c}{(0.866 $\mu$m)} & \multicolumn{1}{c}{(1.57 $\mu$m)} & \multicolumn{1}{c}{(K)} & \multicolumn{1}{c}{($R_{\sun}$)} & \multicolumn{1}{c}{(log $L/L_{\sun}$)} & \\ 
\hline

J22532-142 &GJ 876        	 &M4.0 V&1.815$\pm$0.251&1.847$\pm$0.532&1.469$\pm$0.193&3202$\pm$107&0.275$\pm$0.067&\textendash2.193$\pm$0.225& \\
J22559+178 &GJ 4306       	 &M1.0 V&3.530$\pm$0.165&2.827$\pm$0.391&4.380$\pm$0.272&3750$\pm$120&0.534$\pm$0.070&\textendash1.283$\pm$0.237& \\
J22565+165 &GJ 880        	 &M1.5 V&3.153$\pm$0.115&2.672$\pm$0.273&4.547$\pm$0.141&3703$\pm$112&0.528$\pm$0.068&\textendash1.290$\pm$0.228& \\
J23245+578 &GJ 895        	 &M1.0 V&3.320$\pm$0.147&2.820$\pm$0.384&4.763$\pm$0.091&3751$\pm$107&0.547$\pm$0.067&\textendash1.240$\pm$0.223& \\
J23340+001 &GJ 899        	 &M2.5 V&3.363$\pm$0.223&2.469$\pm$0.525&3.229$\pm$0.224&3629$\pm$111&0.462$\pm$0.068&\textendash1.488$\pm$0.226& \\
J23351-023 &GJ 1286       	 &M5.5 V&0.305$\pm$0.351&1.053$\pm$0.855&0.587$\pm$0.253&2849$\pm$114&0.141$\pm$0.068&\textendash3.004$\pm$0.229& \\
J23381-162 &GJ 4352       	 &M2.0 V&3.193$\pm$0.231&2.790$\pm$0.595&2.205$\pm$0.141&3533$\pm$109&0.392$\pm$0.067&\textendash1.745$\pm$0.225& \\
J23419+441 &GJ 905        	 &M5.0 V&0.456$\pm$0.298&0.679$\pm$0.807&0.879$\pm$0.103&2904$\pm$125&0.171$\pm$0.069&\textendash2.696$\pm$0.231& \\
J23431+365 &GJ 1289       	 &M4.0 V&1.876$\pm$0.242&1.051$\pm$0.624&0.990$\pm$0.251&3192$\pm$112&0.274$\pm$0.068&\textendash2.065$\pm$0.227& \\
J23492+024 &GJ 908        	 &M1.0 V&2.715$\pm$0.091&2.057$\pm$0.264&2.853$\pm$0.076&3475$\pm$104&0.407$\pm$0.066&\textendash1.652$\pm$0.221& \\
J23505-095 &GJ 4367       	 &M4.0 V&1.636$\pm$0.397&1.580$\pm$0.887&1.835$\pm$0.359&3191$\pm$128&0.283$\pm$0.070&\textendash2.132$\pm$0.234& \\
J23556-061 &GJ 912        	 &M2.5 V&3.230$\pm$0.120&2.580$\pm$0.316&3.705$\pm$0.300&3639$\pm$104&0.478$\pm$0.067&\textendash1.447$\pm$0.223& \\
J23585+076 &GJ 4383       	 &M3.0 V&2.569$\pm$0.247&2.279$\pm$0.559&1.670$\pm$0.167&3373$\pm$114&0.333$\pm$0.068&\textendash1.939$\pm$0.231& \\

\hline
\end{tabular}
 \vspace{1ex} 

 \small \textbf{Notes.}
     \raggedright \small $^{(a)}$ Karmn ID are from Reiners et al. (\citet{Reiners2018}). \raggedright \small $^{(b)}$ Spectral types are from Alonso-Floriano
et al. (\citet{Alonso2015}) and references therein. \raggedright \small $^{(c)}$ $T_{\rm eff}$ , radius and luminosity are estimated in this work using equations (\ref{eq2}), (\ref{eq3}) and (\ref{eq4}) respectively. \\     
\end{table*}

\label{lastpage}

\end{document}